\documentclass[useAMS,usenatbib]{mnras}    

\usepackage[caption=false]{subfig}
\usepackage[final]{graphicx}
\usepackage{epstopdf}
\usepackage{mathptmx}
\usepackage{float}
\usepackage{fixltx2e}
\usepackage{natbib}
\usepackage{amsmath}
\usepackage{amssymb}
\usepackage[usenames,dvipsnames]{color}
\usepackage{appendix}
\usepackage[mathscr]{euscript}
\usepackage{afterpage}

\graphicspath{{figures/}}

\newcommand{\mvir}{M_{\text{200c}}}
\newcommand{\rvir}{r_{200c}}

\newcommand{\msun}{\mathrm{M}_\odot}
\newcommand{\mstar}{M_\text{star}}

\newcommand{\hi}{\textsc{\mbox{H\hspace{1pt}i}}}

\newcommand{\eagle}{EAGLE}

\title[Hydrangea]{The Hydrangea simulations: galaxy formation in and around massive clusters}
\author[Y.~Bah\'{e} et al.]{\parbox[t]{\textwidth}{
Yannick M.~Bah\'{e}$^{1}$\thanks{\href{mailto:ybahe@mpa-garching.mpg.de}{ybahe@mpa-garching.mpg.de}}, 
David~J.~Barnes$^{2}$, 
Claudio Dalla Vecchia$^{3, 4}$, Scott T.~Kay$^{2}$, Simon D.~M.~White$^{1}$, Ian G.~McCarthy$^{5}$, Joop Schaye$^{6}$, Richard G.~Bower$^{7}$, Robert A.~Crain$^{5}$, Tom Theuns$^{7}$, Adrian Jenkins$^{7}$, Sean L.~McGee$^{8}$, Matthieu Schaller$^{7}$, Peter A.~Thomas$^{9}$, and James W.~Trayford$^{7}$\vspace*{12pt}}
\\
$^1$ Max-Planck-Institut f\"{u}r Astrophysik, Karl-Schwarzschild Str. 1, 85748 Garching, Germany\\
$^2$ Jodrell Bank Centre for Astrophysics, School of Physics and Astronomy, The University of Manchester, Manchester M13 9PL, UK\\
$^3$ Instituto de Astrof\'{i}sica de Canarias, C/V\'{i}a L\'{a}ctea s/n, E-38205 La Laguna, Tenerife, Spain\\
$^4$ Departamento de Astrof\'{i}sica, Universidad de La Laguna, Av. del Astrof\'{i}sico Francisco S\'{a}nchez s/n, E-38206 La Laguna, Tenerife, Spain\\
$^5$ Astrophysics Research Institute, Liverpool John Moores University, 146 Brownlow Hill, Liverpool, L3 5RF, UK\\
$^6$ Leiden Observatory, Leiden University, PO Box 9513, 2300 RA Leiden, The Netherlands\\
$^7$ Institute for Computational Cosmology, Department of Physics, University of Durham, South Road, Durham DH1 3LE, UK\\
$^8$ School of Physics and Astronomy, University of Birmingham, Edgbaston, Birmingham, B15 2TT, UK\\
$^9$ Astronomy Centre, University of Sussex, Falmer, Brighton BN1 9QH, UK\\
}

\begin{document}
\label{firstpage}
\maketitle

\begin{abstract}
We introduce the Hydrangea simulations, a suite of 24 cosmological hydrodynamic zoom-in simulations of massive galaxy clusters ($\mvir = 10^{14}$--$10^{15.4}\, \msun$) with baryon particle masses of $\sim$$10^6\, \msun$. Designed to study the impact of the cluster environment on galaxy formation, they are a key part of the `Cluster-EAGLE' project (Barnes et al., in prep.). They use a galaxy formation model developed for the \eagle{} project, which has been shown to yield both realistic field galaxies and hot gas fractions of galaxy groups consistent with observations. The total stellar mass content of the simulated clusters agrees with observations, but central cluster galaxies are too massive, by up to 0.6 dex. Passive satellite fractions are higher than in the field, and at stellar masses $\mstar > 10^{10}\, \msun$ this environmental effect is quantitatively consistent with observations. The predicted satellite stellar mass function matches data from local cluster surveys. Normalized to total mass, there are fewer low-mass ($\mstar \lesssim 10^{10}\, \msun$) galaxies within the virial radius of clusters than in the field, primarily due to star formation quenching. Conversely, the simulations predict an overabundance of massive galaxies in clusters compared to the field that persists to their far outskirts ($> 5\,\rvir$). This is caused by a significantly increased stellar mass fraction of (sub-)haloes in the cluster environment, by up to $\sim$0.3 dex even well beyond $\rvir$. Haloes near clusters are also more concentrated than equally massive field haloes, but these two effects are largely uncorrelated.
\end{abstract}



\section{Introduction}
\label{sec:introduction}

In the local Universe, strong correlations exist between the properties of galaxies and their large-scale environment. In particular, galaxies in groups and clusters are typically red, lack recent and ongoing star formation (e.g.~\citealt{Balogh_et_al_1999, Kauffmann_et_al_2004, Weinmann_et_al_2006, Peng_et_al_2010, Wetzel_et_al_2012}), are depleted in atomic hydrogen (\hi; \citealt{Giovanelli_Haynes_1985, Fabello_et_al_2012, Hess_Wilcots_2013}), and biased towards early-type (elliptical) morphologies (e.g.~\citealt{Dressler_1980}). 

However, all of these properties are also observed to correlate with galaxy luminosity and stellar mass, so that it is possible that these differences stem, at least in part, from different stellar mass distributions between dense environments and the field. The luminosity function of cluster galaxies has been studied by several authors in the last decade (e.g.~\citealt{Popesso_et_al_2006, Agulli_et_al_2014, Agulli_et_al_2016, Lan_et_al_2016}). Some of these works indeed found significant variations of the luminosity function between clusters and the field, especially in the form of a steep faint-end upturn in clusters (\citealt{Popesso_et_al_2006, Lan_et_al_2016}). However, the deep observations of the cluster Abell 85 by \citet{Agulli_et_al_2014, Agulli_et_al_2016} found no evidence for such a steep upturn. This uncertainty complicates the interpretation of the observed environmental variations of galaxy properties. 

Stellar mass is arguably a more fundamental quantity than luminosity, but its determination requires estimating the mass-to-light ratio from galaxy colours (e.g.~\citealt{Bell_deJong_2001}), or, if available, spectra (e.g.~\citealt{Kauffmann_et_al_2003a, Gallazzi_et_al_2005}). From an analysis of SDSS spectroscopic data, \citet{Kauffmann_et_al_2004} demonstrated that a larger fraction of stellar mass in dense environments is contributed by more massive galaxies compared to low-density regions. Subsequent studies have suggested that this shift is driven mainly by the special properties of central cluster galaxies (e.g. \citealt{vonDerLinden_et_al_2010}): \citet{Calvi_et_al_2013}, for example, report that the shape of the \emph{satellite} stellar mass function in clusters is similar to that in the field, at least at the massive end. Several other authors, however, have found differences between the satellite and field stellar mass functions, at either the high- or low-mass end \citep{Yang_et_al_2009, Wang_White_2012, Vulcani_et_al_2014}. In part, these differences may be driven by different definitions of `environment' (local density, halo mass, radial range) and differences in accounting for fore-/background galaxies.

An observational consensus on the nature of stellar mass differences in different environments would clearly be desirable, but even in its absence one can gain valuable insight into the expected extent of, and physical reason underlying, such differences through predictions from theoretical galaxy formation models. Cosmological hydrodynamic simulations are able to self-consistently predict differences in the formation of central and satellite galaxies, without explicitly prescribing the action of specific processes affecting only the latter. This gives them, in principle, great predictive power to understand the star formation histories of cluster galaxies as manifested in their present-day stellar masses. 

However, such simulations have for a long time been unable to predict a galaxy stellar mass function in the field that agrees with observations (e.g.~\citealt{Crain_et_al_2009, Scannapieco_et_al_2012}), which is clearly a prerequisite for making meaningful predictions about galaxies in clusters. This problem has been solved only recently, thanks to increased resolution and, in particular, significant efforts to improve and calibrate the subgrid models that the simulations employ to model the unresolved aspects of feedback from star formation and accreting supermassive back holes. With these improvements, the \eagle{} \citep{Schaye_et_al_2015, Crain_et_al_2015} project has produced a simulation that could be calibrated to match the observed stellar mass function and sizes of present-day field galaxies (see also \citealt{Vogelsberger_et_al_2014} and \citealt{Dubois_et_al_2014} for the similarly successful Illustris and Horizon-AGN projects). Apart from these calibrated matches, \eagle{} has also successfully reproduced, amongst others, the observed colour bimodality of galaxies \citep{Trayford_et_al_2015}, the evolution of galaxy sizes and star formation rates \citep{Furlong_et_al_2015, Furlong_et_al_2017}, their black hole mass function \citep{Rosas-Guevara_et_al_2016} and the correlation between galactic star formation and black hole accretion rates \citep{McAlpine_et_al_2017}, their atomic \citep{Rahmati_et_al_2015, Bahe_et_al_2016, Crain_et_al_2017} and molecular hydrogen content \citep{Lagos_et_al_2015}, and the environmental effect of galaxy groups on atomic hydrogen \citep{Marasco_et_al_2016} and galaxy metallicity \citep{Bahe_et_al_2017}. 

Galaxy clusters, however, occupy only a small volume fraction of the Universe, so that simulation volumes much larger than available in \eagle{} are necessary to sample them in representative numbers. Such simulations can, at present, only afford a much lower resolution of $\gtrsim$ 5 kpc in spatial terms or particle masses of $m_\text{baryon} \approx 10^9\, \msun$ (e.g.~\citealt{LeBrun_et_al_2014,Bocquet_et_al_2016,McCarthy_et_al_2016}), compared to 0.7 kpc and $\sim$10$^6\, \msun$ for \eagle{}. This precludes studying even basic predictions such as stellar masses for galaxies with $\mstar \lesssim 10^{10}\, \msun$, while more numerically sensitive properties such as their atomic gas content or metallicity are inaccessible for all but the most massive galaxies.
 
Until simulations at the resolution of \eagle{}, but with orders-of-magnitude larger volume, become computationally feasible, progress can still be made through zoom-in simulations, where only a small, carefully selected volume inside a much larger parent simulation is modelled at high resolution and including baryons. The bulk of the volume is instead filled with low-resolution boundary particles interacting only through gravity, whose purpose is the creation of appropriate tidal fields and large-scale modes in the high-resolution region (e.g.~\citealt{Katz_White_1993, Tormen_et_al_1997, Borgani_et_al_2002, Dolag_et_al_2009, Martizzi_et_al_2014, Hahn_et_al_2015, Barnes_et_al_2017}). 

Motivated by these considerations, this paper introduces the \emph{Hydrangea} simulation project\footnote{Named after the plant \emph{Hydrangea macrophylla}, whose petals change their colour from blue to red according to their environment, in analogy to the colour--density relation of galaxies.}, a suite of 24 high-resolution zoom-in galaxy clusters run with the \eagle{} code for the purpose of studying the interaction between clusters and the galaxies in and around them. Each high-resolution simulation region is centred on a massive cluster ($\mvir = 10^{14.0}-10^{15.4} \msun$)\footnote{$\mvir$ denotes the total mass within a sphere of radius $\rvir$, centred on the potential minimum of the cluster, within which the average density equals 200 times the critical density.}, and realized at the same resolution level as the largest-volume simulation of the \eagle{} project ($m_\text{baryon} = 1.81 \times 10^6\, \msun$, gravitational softening length $\epsilon = 0.7$ physical kpc at $z < 2.8$). The high-resolution zoom-in region is set up to include not only the cluster haloes themselves, but also their large-scale surroundings out to ten virial radii, i.e.~$\sim$10--25 comoving Mpc, motivated by indications from observations (e.g.~\citealt{vonDerLinden_et_al_2010, Wetzel_et_al_2012, Lu_et_al_2015}) and theory \citep{Bahe_et_al_2013} that the environmental influence on at least some galaxy properties extends significantly beyond the virial radius. 

In this paper, we present a validation of the simulations in terms of some of the most fundamental galaxy properties, namely their stellar mass function and quenched fractions at $z \approx 0$, and then use the detailed information provided by the simulations to gain insight into the impact of the cluster environment on the galaxy stellar mass function. In a companion paper (Barnes et al., in prep.), we analyze the properties of the hot intracluster medium in a sample of simulated clusters including the Hydrangea suite, and demonstrate that the simulations predict X-ray and SZ properties that are broadly compatible with low-redshift observational constraints. Predictions for the galaxy luminosity functions in our simulations, including results from a higher-resolution run of an intermediate-mass cluster, will be presented by Dalla Vecchia et al. (in prep.). Together, these simulations form the `C-EAGLE' project family\footnote{`Cluster-EAGLE', also referring to Steller's sea eagle (\emph{Haliaeetus pelagicus}) as the largest member of the avian eagle family.}.

This paper is structured as follows. In \S \ref{sec:sims}, we review the \eagle{} galaxy formation model that was used in our simulations, and describe the selection and simulation of the clusters that form the Hydrangea suite. We then compare several key predictions of the simulations to $z \approx 0$ observations in \S \ref{sec:obscomp}, followed by a detailed analysis of the simulated stellar mass function in \S \ref{sec:gsmf}. Our results are then summarised and discussed in \S \ref{sec:summary}.

Throughout the paper, we use the same flat $\Lambda$CDM cosmology as used in the \eagle{} simulations, with parameters as determined by \citet{Planck_2014}: Hubble parameter $h \equiv $ H$_{0}/(100\,{\rm km}\,{\rm s}^{-1}{\rm Mpc}^{-1}) = 0.6777$, dark energy density parameter $\Omega_\Lambda = 0.693$ (dark energy equation of state parameter $w=-1$), matter density parameter $\Omega_{\rm M} = 0.307$, and baryon density parameter $\Omega_{\rm b} = 0.04825$. For length scales, the prefix `p' and `c' denotes physical and comoving quantities, respectively (e.g. `pkpc' for `physical kpc'); where no prefix is given, distances are given in physical units. Unless otherwise specified, all galaxy masses are computed as the sum of gravitationally bound star particles within 30 pkpc from the potential minimum of their subhalo (see \citealt{Schaye_et_al_2015}).


\section{Description of the simulations}
\label{sec:sims}

In this Section, we first provide a summary of the key features of the \eagle{} code that was used for this work (\S \ref{sec:eagle}), and then describe the setup and running of the Hydrangea cluster simulations (\S \ref{sec:hydrangea}).

\subsection{The \eagle{} galaxy formation model}
\label{sec:eagle}

The simulation code developed for the \eagle{} project is a substantially modified version of the \textsc{gadget}-3 smoothed particle hydrodynamics (SPH) code, last described in \citet{Springel_et_al_2005}. We restrict our description here to a summary of only its key features and refer the interested reader to the detailed description by \citet{Schaye_et_al_2015}.

Compared to \textsc{gadget-3}, the hydrodynamics and timestepping scheme has undergone several updates that are collectively referred to as ``Anarchy'' (Dalla Vecchia, in prep.; see also Appendix A of \citealt{Schaye_et_al_2015} and \citealt{Schaller_et_al_2015}). These include using the conservative pressure-entropy formulation of SPH \citep{Hopkins_2013}, an artificial viscosity switch \citep{Cullen_Dehnen_2010}, an artificial conduction switch similar to that of \citet{Price_2008}, the $C^2$ \citet{Wendland_1995} kernel, and the time-step limiter proposed by \citet{Durier_DallaVecchia_2012}.  These updates mitigate many of the shortcomings of `traditional' SPH codes, such as the treatment of surface discontinuities, described by e.g.~\citet{Agertz_et_al_2007} and \citet{Mitchell_et_al_2009}. \citet{Schaller_et_al_2015} discuss the impact of these modifications on the simulated galaxies in detail, and show that the most significant change is due to the \citet{Durier_DallaVecchia_2012} time-step limiter. These authors also demonstrated that the improved hydrodynamics implementation is a key requirement for the efficient action of feedback from supermassive black holes, as described further below. 

Most importantly, the code contains subgrid physics models that were evolved from those developed for the OWLS \citep{Schaye_et_al_2010} simulation project. 

Radiative cooling and photoheating rates are computed on an element-by-element basis following \citet{Wiersma_et_al_2009a}, by considering the 11 most important atomic coolants (H, He, C, N, O, Ne, Mg, Si, S, Ca, Fe) in ionization equilibrium and in the presence of a \citet{Haardt_Madau_2001} ionizing UV/X-ray background. As discussed by \citet{Schaye_et_al_2015}, the code does not account for self-shielding of gas, because in the regime where this is expected to be important ($n_\mathrm{H} \gtrsim 10^{-2}\, \text{cm}^{-2}$), the uncertain effect of local stellar radiation would also need to be considered \citep{Rahmati_et_al_2013b}. 

The modelling of reionization follows \citet{Wiersma_et_al_2009b}. To account for hydrogen reionization, the \citet{Haardt_Madau_2001} background is switched on at redshift $z = 11.5$ \citep{Theuns_et_al_2002a, Planck_2014p1}. This is accompanied by the injection of 2 eV of energy per proton mass. He reionization is modelled by injecting the same amount of energy around $z = 3.5$, which results in a thermal evolution of the IGM in agreement with the observations of \citet[see also \citealt{Theuns_et_al_2002b}]{Schaye_et_al_2000}.

The star formation rate of gas particles is modelled as a pressure-law following \citet{Schaye_DallaVecchia_2008}, 
\begin{equation}
\dot{m}_\text{star} = m_\text{g} A \left( 1\, \msun\, \text{pc}^{-2} \right)^{-n} \left(\frac{\gamma}{G} P \right)^{(n-1)/2}, 
\label{eq:sfr}
\end{equation}
where $\dot{m}_\text{star}$ is the star formation rate of a gas particle with mass $m_g$ and (total) pressure $P$, $\gamma = 5/3$ is the ratio of specific heats, and $G$ the gravitational constant. The subgrid parameters $A = 1.515 \times 10^{-4}\, \msun\, \text{yr}^{-1}\, \text{kpc}^{-2}$ and $n = 1.4$ are then directly prescribed by observations \citep{Kennicutt_1998}, independent of any imposed equation of state. Deviating from \citet{Schaye_DallaVecchia_2008}, the star formation threshold $n_\mathrm{H}^*$ depends on metallicity, as proposed by \citet{Schaye_2004}:
\begin{equation}
n_\mathrm{H}^*(Z) = 10^{-1} \text{cm}^{-3} \left(\frac{Z}{0.002} \right)^{-0.64}, 
\end{equation}
where $Z$ is the gas-phase metallicity smoothed over the SPH kernel (see \citealt{Wiersma_et_al_2009b}). This equation accounts for the metallicity dependence of the transition from the warm atomic to the cold molecular interstellar gas phase. $n_\mathrm{H}^*(Z)$ is limited to a maximum of 10 cm$^{-3}$ to prevent divergence at low $Z$. Star formation is then implemented stochastically with the probability of a gas particle being converted to a star set by equation \eqref{eq:sfr}. Because the simulations lack the resolution and physics to directly model the cold dense gas phase in which star formation is observed to occur in the real Universe, a pressure floor corresponding to $P_\text{eos} \propto \rho_g^{4/3}$ is imposed on gas with $n_\text{H} \geq  10^{-1} \text{cm}^{-3}$, normalized to $T_\text{eos} = 8 \times 10^3\, \text{K}$ at that density. As this relation corresponds to a constant Jeans mass, it prevents artificial fragmentation due to a lack of numerical resolution \citep{Schaye_DallaVecchia_2008}.

Mass and metal enrichment of gas due to stellar mass loss is modelled as described by \citet{Wiersma_et_al_2009b} with the modifications described in \citet{Schaye_et_al_2015}. This approach is based on treating star particles as simple stellar populations with a \citet{Chabrier_2003} IMF in the mass range 0.1--100 $\msun$ and accounting for winds from AGB and massive stars as well as type-Ia and core-collapse supernovae.

Energy feedback from star formation is implemented in a single thermal mode, by heating a small number of gas particles ($\sim$1) by a large temperature increment ($\Delta T = 10^{7.5} \text{K}$). \citet{DallaVecchia_Schaye_2012} demonstrate that this approach alleviates numerical overcooling without the need to temporarily disable hydrodynamic forces or radiative cooling for affected gas particles, but can still not avoid it completely in the regions where the gas density is highest, and the cooling time therefore shortest. As discussed in detail by \citet{Crain_et_al_2015}, the efficiency of star formation feedback is therefore scaled with gas density so that energy input in dense regions formally exceeds the physically available energy budget from core-collapse supernovae. Averaged over the entire simulation, however, the ratio is below unity. In addition, the efficiency is lowered in high-metallicity gas to account for the physically expected higher cooling losses. \citet{Crain_et_al_2015} show that these scalings of star formation feedback efficiency are crucial for obtaining galaxies with realistic sizes, although the total galaxy masses are largely insensitive to them.

We note that, as an undesired side effect, these high-energy, stochastic, local heating events produce gas discs in some simulated galaxies that contain artificially large holes \citep{Bahe_et_al_2016}. As we discuss further in Section \ref{sec:sfr}, these holes may affect the predicted interaction between the dense cold gas discs and the hot intra-cluster gas in our simulations.

Finally, the code includes a model for the growth of supermassive black holes (BHs), which are seeded in a friends-of-friends (FoF) halo once its mass exceeds $10^{10} \,h^{-1} \msun$ \citep{Springel_et_al_2005b} with a (subgrid) black hole seed mass of $10^5\, h^{-1} \msun$. Subsequently, the subgrid BH mass grows as a consequence of gas accretion, which is modelled as in \citet{Rosas-Guevara_et_al_2015} but without the \citet{Booth_Schaye_2009} `boost factor' \citep{Schaye_et_al_2015}. In essence, this approach considers the angular momentum of gas near the black hole to limit the Bondi accretion rate to 
\begin{equation}
\dot{m}_\text{accr} = \dot{m}_\text{Bondi} \times \text{min} \left(C_\text{visc}^{-1} \left(c_s/V_\phi\right)^3, 1\right)
\end{equation}
where $c_s$ is the sound speed and $V_\phi$ the rotation speed of gas around the black hole. The parameter $C_\text{visc}$ was thought to set the stellar mass at which accretion becomes efficient \citep{Rosas-Guevara_et_al_2015}\footnote{However, \citet{Bower_et_al_2017} have shown that this scale is instead determined by the critical halo mass above which the hot hydrostatic atmosphere traps outflows driven by star formation and is nearly independent of $C_\text{visc}$.}. In the \eagle{} reference model (`Ref'), $C_\text{visc} = 2\upi$. 

In analogy to star formation, energy feedback from supermassive black holes (`AGN feedback') is implemented stochastically, with one particle heated by a large temperature increment. Following \citet{Booth_Schaye_2009}, 15 per cent of the accreted rest mass is converted to energy, with a 10 per cent coupling efficiency to the surrounding gas, i.e. an energy injection rate of $0.015\, \dot{m}_\text{accr}\, c^2$ (where $c$ is the speed of light). Because the gas surrounding supermassive black holes is typically denser than around newly formed stars, the temperature increment $\Delta T_\mathrm{AGN}$ must also be higher to make the feedback efficient. In the Ref model, one particle per heating event is heated by $\Delta T_\mathrm{AGN} = 10^{8.5} \mathrm{K}$. However, \citet{Schaye_et_al_2015} have shown that this predicts X-ray luminosities and hot gas fractions in galaxy groups and intermediate-mass clusters that are higher than observed. An alternative model that differs from Ref only in its choice of $\Delta T_\mathrm{AGN}\, (=10^9 \mathrm{K})$ and $C_\mathrm{visc}\, (=2\upi \times 10^2)$, `AGNdT9', was shown to largely resolve these discrepancies on the scale of galaxy groups, while achieving a similarly good match as Ref to observed properties on galactic scales\footnote{Because AGNdT9 was only realized in a (50 cMpc)$^3$ simulation volume, it contains only one halo whose mass at $z=0$ is (just) above $10^{14}\, \msun$. \citet{Schaye_et_al_2015} could therefore not test its predictions on the hot gas properties in massive clusters.}. We therefore adopt the AGN feedback parameterisation of AGNdT9 for all C-EAGLE simulations, including the Hydrangea suite presented here. In a companion paper (Barnes et al., in prep.), we show that this model also leads to simulated \emph{clusters} with overall realistic intra-cluster medium (ICM) properties, albeit with a still somewhat too high hot gas mass fraction (by $\sim2\sigma$), and artificially high entropy levels in the cluster cores.

\subsection{The Hydrangea simulations}
\label{sec:hydrangea}

\subsubsection{Selection of the C-EAGLE cluster sample}

The reason for the absence of massive galaxy clusters in the original \eagle{} simulations is their relatively small volume of $\leq (100 \,\text{cMpc})^3$. Our new simulations are therefore based on a much larger `parent simulation', described by \citet{Barnes_et_al_2017}. This is a (3200 cMpc)$^3$ volume which was simulated with dark matter only, in the same cosmology as that adopted for the \eagle{} project \citep[see Introduction]{Planck_2014}. The dark matter particle mass in the parent simulation is 8.01 $\times 10^{10}\, \msun$ with a gravitational softening length of 59 ckpc; a galaxy cluster with $M > 10^{14}\, \msun$ is therefore resolved by at least 1000 particles.

From the parent simulation snapshot at $z=0$, we then selected candidate clusters for zoom-in re-simulation. Apart from a threshold in halo mass ($\mvir \geq 10^{14}\, \msun$), we also applied a mild isolation criterion, by requiring that no more massive halo be located within 30 pMpc, or 20 $\rvir$, whichever is larger ($\rvir$ here refers to the radius of the neighbouring, more massive halo), from any re-simulation candidate. This criterion ensures that our simulations are centred on the peak of the local density structure and not, for example, on a moderately massive halo on the outskirts of an even more massive cluster. Finally, for computational convenience we required that our candidate clusters be no closer than 200 pMpc to any of the periodic simulation box edges. 

From this initial list of 91,824 candidate haloes, we then selected a subset of 30 objects for re-simulation. To avoid a bias towards the more common lower-mass haloes, our candidates were binned by $\mvir$ into ten logarithmic bins from $10^{14}\, \msun$ to $2\times10^{15}\, \msun$ ($\Delta \log_{10} \mvir$ = 0.13). Three objects were then selected from each bin at random. To extend our mass range yet further, we only picked two objects from the highest mass bin, and selected a final halo at even higher mass, $\mvir = 10^{15.34}\, \msun$. These thirty objects comprise the C-EAGLE cluster sample.

\subsubsection{Motivation for large zoom-in regions}
\label{sec:large_motivation}
The virial radius, approximated by $\rvir$, has traditionally been assumed to represent the boundary between a halo and the surrounding Universe, based on the spherical collapse model. However, evidence has emerged in recent years that galaxies might be affected by their environment out to significantly larger distances (e.g.~\citealt{Balogh_et_al_1999, Haines_et_al_2009, Hansen_et_al_2009, vonDerLinden_et_al_2010, Lu_et_al_2012, Rasmussen_et_al_2012, Wetzel_et_al_2012}), a result that has been supported by previous generation hydrodynamical simulations \citep{Bahe_et_al_2013, Bahe_McCarthy_2015}. While most observational evidence for this large-scale influence is based on galaxy colours and star formation rates, \citet{Bahe_et_al_2013} have shown that the GIMIC simulations predict an effect that reaches even further when the hot gas haloes of galaxies are considered instead: in galaxies with $\mstar \approx 10^9\, \msun$, these are predicted to be depleted even at $r > 5\,\rvir$ from the centre of a group or cluster. 

Simulations aiming to shed light onto the mechanisms affecting galaxy evolution in dense environments should therefore not be limited to the dense cluster haloes alone (within $\sim$$\rvir$), but also extend far enough into the surrounding volume to capture the large-scale environmental impact. The disadvantage of this is a significant increase of the high-resolution simulation volume, increasing both computing time and especially the memory footprint of the simulation. To strike a balance between these conflicting constraints, we simulated 24 of the 30 C-EAGLE clusters with zoom-in regions extending to at least $10\, \rvir$ from the cluster centre; these objects constitute the Hydrangea simulations as analysed in this paper. The remaining six objects, with masses between $10^{14.6}$ and $10^{15.2}\, \msun$, were simulated only out to $5\,\rvir$, primarily serving as tools to study the ICM for which each simulation only contributes one (central) object of interest, as opposed to several hundreds or even thousands of galaxies. The additional C-EAGLE simulations are described in more detail by Barnes et al. (in prep.). 

\subsubsection{Simulation runs and post-processing}
\label{sec:runs}

The Hydrangea simulations were run mostly on the HazelHen Cray-XC40 system hosted by the German Federal Maximum Performance Computing Centre (HLRS) at the University of Stuttgart. This system provides nodes with 128 GB of memory each, shared by 24 compute cores for an effectively available 5 GB of memory per core. On this system, we could accommodate most of our hydrodynamic runs on $\leq 2048$ cores to minimize scaling losses in our highly clustered simulations. From initial conditions generated as described in Appendix \ref{app:ics} (see also Barnes et al., in prep.), the most massive cluster in our sample required more than 10 million core hours to reach $z = 0$, corresponding to a total wall clock time of over ten months (including queueing and downtime). Several clusters from the low-mass end of our sample were run on machines at the Max Planck Computing and Data Facility (MPCDF) in Garching. 

In addition to these hydrodynamic simulations we also performed one DM-only simulation of each zoom-in region. These use the same initial conditions as the hydrodynamical runs, but due to their non-dissipative nature, they produce less small-scale clustering and hence only consumed $<\, 10^5$ CPU-hrs each.

As main output from the simulations, 30 full `snapshots' were stored between $z = 14.0$ and $z = 0$. Out of these, 28 are spaced equidistant in time ($\Delta t = 500\,\, \text{Myr}$), while two additional snapshots (at $z = 0.101$ and $z = 0.366$) were included to facilitate comparison to the \eagle{} simulations\footnote{Including these two extra snapshots, 12 \eagle{} snapshots have a counterpart in Hydrangea with a time offset of $\loa$ 50 Myr, including eight at $z \loa 2.0$.}. All snapshots were post-processed with the \textsc{subfind} code \citep{Springel_et_al_2001b,Dolag_et_al_2009} to identify friends-of-friends (FoF) haloes, using a linking length of $b = 0.2$ times the mean inter-particle separation, and self-bound subhaloes within them. We note in this context that `subhalo' can refer to either the central object that contains the largest fraction of the FoF mass or (where they exist) less massive `satellites'.  

Subhaloes in the DM-only and hydrodynamic runs were individually matched by comparing their unique DM particle IDs, as described by \citet{Velliscig_et_al_2014} and \citet{Schaller_et_al_2015A}. The fifty most-bound DM particles in each subhalo from the DM-only simulation are located in the corresponding hydrodynamic simulation. If one subhalo contains at least half of the particles with the same ID in the hydrodynamic simulation, a link is initiated between the two. This link is then confirmed if, and only if, the original subhalo in the DM-only simulation also contains at least 25 of the 50 most-bound DM particles of the corresponding subhalo in the hydrodynamic simulation. 92 per cent of central subhaloes with $\mvir > 10^{11} \msun$ could be successfully matched between the hydrodynamic and DMO simulations in this way.

To reconstruct the evolutionary and orbital histories of individual simulated galaxies, we have linked subhaloes between different snapshots using an updated version of the algorithm described in \citet{Bahe_McCarthy_2015}. This method is described in full in Appendix \ref{app:tracing}. In essence, subhaloes in adjacent snapshots are linked by matching their constituent DM particles, taking into account the formation of new galaxies, mergers between them, and temporary non-identification of galaxies by the \textsc{subfind} algorithm in dense environments (see e.g.~\citealt{Muldrew_et_al_2011}). We note that this algorithm is similar, but not identical, to that used by \citet{Qu_et_al_2017} to build merger trees from the \eagle{} simulations. Unlike in \citet{Bahe_McCarthy_2015}, we base the tracing on DM particles only. This simplification is possible because of the higher resolution of the Hydrangea simulations, which allows DM haloes to be resolved even for galaxies undergoing severe stripping.

In addition, we stored a larger number of `snipshots' that contain only the most important, and most rapidly time-varying, quantities, such as particle positions and velocities (similar to \eagle{}; see \citealt{Schaye_et_al_2015}). We stored three snipshots between each of the 28 main snapshots, for a combined time resolution of $\Delta t = 125\,\, \text{Myr}$. This was then additionally boosted to $\Delta t = 25 \,\,\text{Myr}$ for three 1-Gyr intervals at lookback times of 0--1, 4--5, and 7--8 Gyr. For one intermediate-mass cluster, snipshots were stored at a constant time interval of $\Delta t = 12.5\,\, \text{Myr}$. In future papers, we will exploit the high time resolution provided by these snipshot outputs to trace the evolution of our simulated cluster galaxies in detail; here, we restrict ourselves to an analysis of the snapshot data, in particular those at $z=0$ and $z=0.101$.

\subsubsection{Visualizations of the simulated clusters}
\begin{figure*}
  \centering
    \includegraphics[width=2\columnwidth]{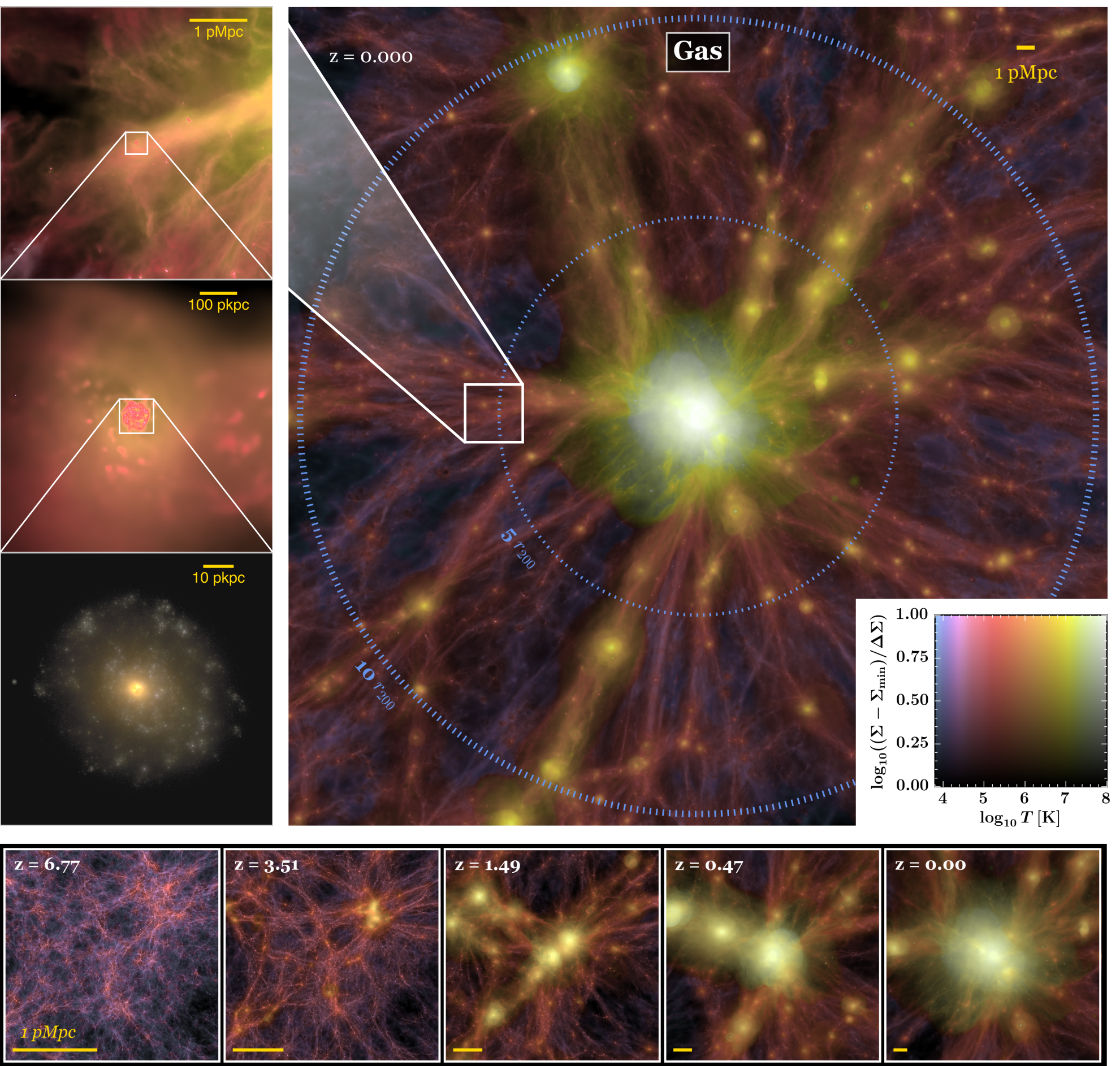}    
       \caption{Visualization of the gas distribution at redshift $z=0$ centred on the most massive Hydrangea cluster (CE-29 with $\mvir = 10^{15.38}\, \msun$). The main panel presents a 60x60x15 pMpc slice centred on the potential minimum of the cluster, with gas surface density and temperature represented, respectively, by the image brightness and hue/saturation (see color map in the bottom-right corner). The two dotted blue rings indicate mid-plane distances of 5 and 10 $\rvir$ from the cluster centre; the latter corresponds to the nominal edge of the high-resolution region. The panels on the left-hand side zoom in towards one individual galaxy on the cluster outskirts, highlighting the detailed internal structure that is resolved in our simulations; the bottom panel shows a synthetic optical \emph{gri} image of the galaxy. The five panels in the bottom row show the gas distributions at different redshifts; each is a projected cube with side length 20 $h^{-1}$ cMpc centred on the main progenitor of the $z=0$ cluster. For reference, a physical length scale of 1 pMpc is indicated by the yellow bar in the bottom-left corner of each panel.}
      \label{fig:pretty}
\end{figure*}

A visualization of one Hydrangea simulation is presented in Fig.~\ref{fig:pretty}; this contains at its centre the most massive cluster, CE-29, with $\mvir = 10^{15.38} \msun$\footnote{Note that there are small differences between the halo masses in the low-resolution parent simulation and high-resolution hydrodynamic zoom-in resimulations, by $< 0.05$ dex. As a convention, we denote individual zoom-in regions, and their central clusters, by the prefix `CE' (for C-EAGLE), followed by their ID number from 0 to 29 (see Table \ref{tab:info}).}. The main panel shows the gas distribution at $z=0$ in a slice of side length 60x60 pMpc and thickness 15 pMpc, centred on the potential minimum of the cluster. The colour map, shown in the bottom-right inset, encodes both the projected gas density (as brightness) and temperature (as hue/saturation); the coldest gas ($T \lesssim 10^4$ K) is shown in blue, and the hottest ($T \gtrsim 10^8$ K) in white. Clearly visible is the central hot ($T \gtrsim 10^7$ K) halo that extends to $\sim$4 $\rvir$, and a myriad of filaments and embedded haloes out to the nominal edge of our high-resolution region at $10\,\rvir$ (thick dotted blue line). 

The three panels on the left-hand side present successive zoom-ins towards one individual galaxy on the cluster outskirts, highlighting the vast dynamic range of the simulation. The top two show the gas density and temperature, using the same temperature scaling as the main panel but with adjusted scaling of the surface density for improved clarity. In the bottom panel, we display a synthetic \emph{gri} optical image created with the radiative transfer code `\textsc{Skirt}' (\citealt{Camps_et_al_2016}; Trayford et al. in prep.).

The five panels in the bottom row illustrate the formation history of the cluster. Each shows a projected cube of side length 20 $h^{-1}$ cMpc, centred on the main progenitor of the $z=0$ cluster. The corresponding physical scale is indicated by the yellow bar in the bottom-left corner of each panel, which indicates a length of 1 pMpc. Starting from a web-like structure at $z \approx 7$, the simulation forms a number of proto-cluster cores by $z=1.5$ which then successively merge to form the present-day cluster. As an aside, we note that the main progenitor at high redshift $(z \gtrsim 1)$ is clearly not the most massive proto-cluster core, but the one that experiences the most rapid growth prior to the final merging phase.   

\begin{figure*}
  \centering
    \includegraphics[width=2.07\columnwidth]{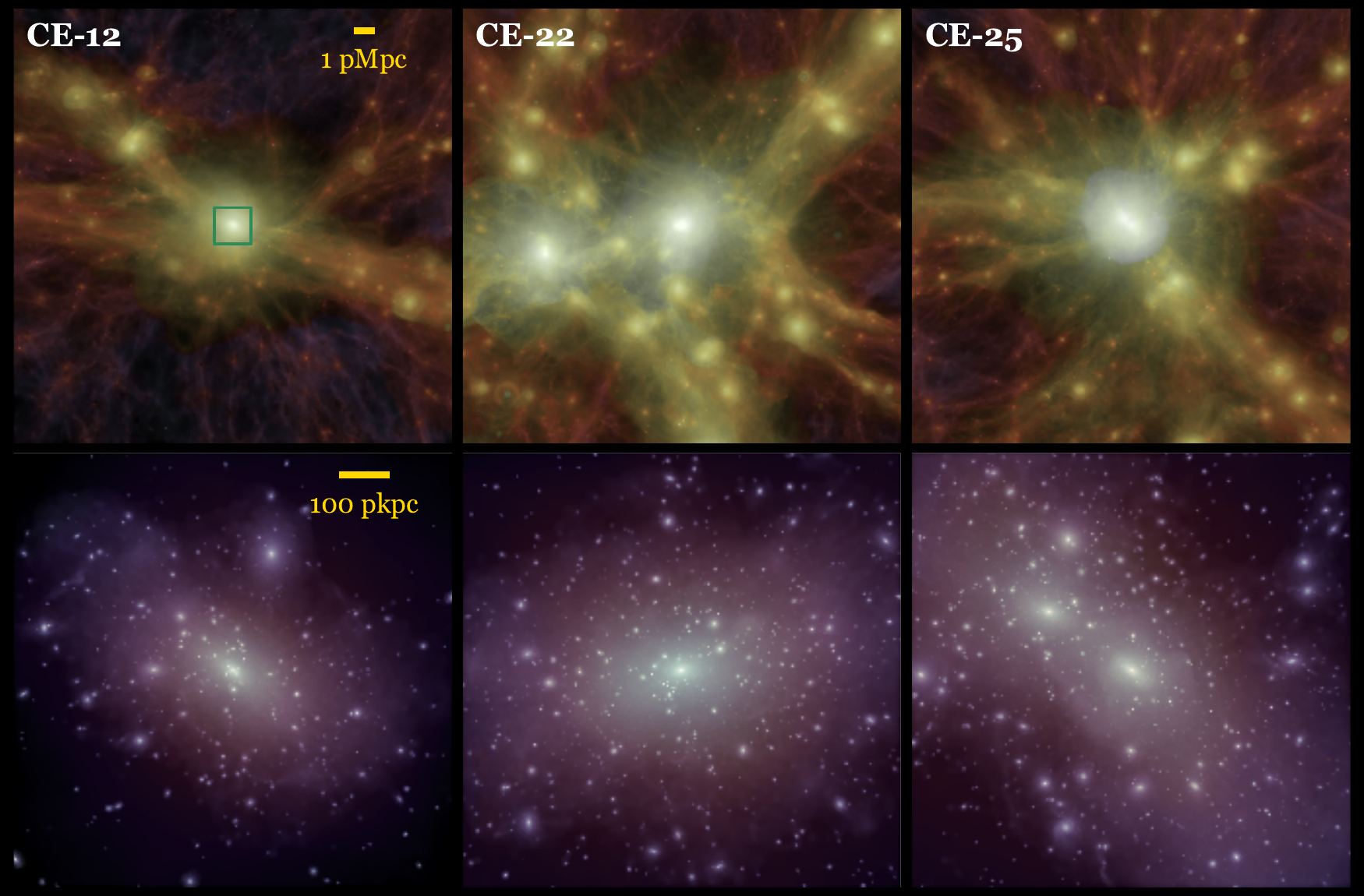}    
       \caption{Three visual examples of the variety of clusters in the mass range $4 \times 10^{14} \msun$ to $1.4 \times 10^{15} \msun$ in the Hydrangea suite at redshift $z=0$. The top row shows the projected gas density in a 30 pMpc cube with colour indicating the gas temperature as in Fig.~\ref{fig:pretty}. The bottom row shows the central 2.5 pMpc of each simulation, with stellar surface density in shades of grey overlaid on gas density (purple through yellow). The green box in the top-left panel indicates the size of the regions depicted on the bottom row.}
      \label{fig:clustercomp}
\end{figure*}

The range of cluster morphologies in our suite, on both large and small scales, is illustrated by Fig.~\ref{fig:clustercomp}. For three clusters, this figure shows the gas density and temperature as in Fig.~\ref{fig:pretty}, projected within a cube of 30 pMpc side length (top row), and in the bottom row the stellar mass surface density (greyscale) blended with the gas density (purple through yellow) within a cube of 2.5 pMpc side length. Both are centred on the potential minimum of the cluster. For guidance, the region depicted in the bottom row is indicated by the green box in the top-left panel. 

The three example clusters are embedded in strikingly different large-scale environments, including a highly isolated object (CE-12, left), a supercluster (CE-22, middle), and a cluster that dominates a region with several less massive haloes (CE-25, right). Similar, but not necessarily correlated, differences are evident in the distribution of galaxies formed from the stars in their centres: some contain a dominating ``brightest cluster galaxy'' (BCG; e.g. CE-12 and CE-22), whereas CE-25 in the right-hand column is currently undergoing a triple-merger without an obvious `central' galaxy\footnote{As can be seen in the top panel, this merger leads to an expansion of the hot halo in a clear shock front.}

\begin{figure}
  \centering
    \includegraphics[width=\columnwidth]{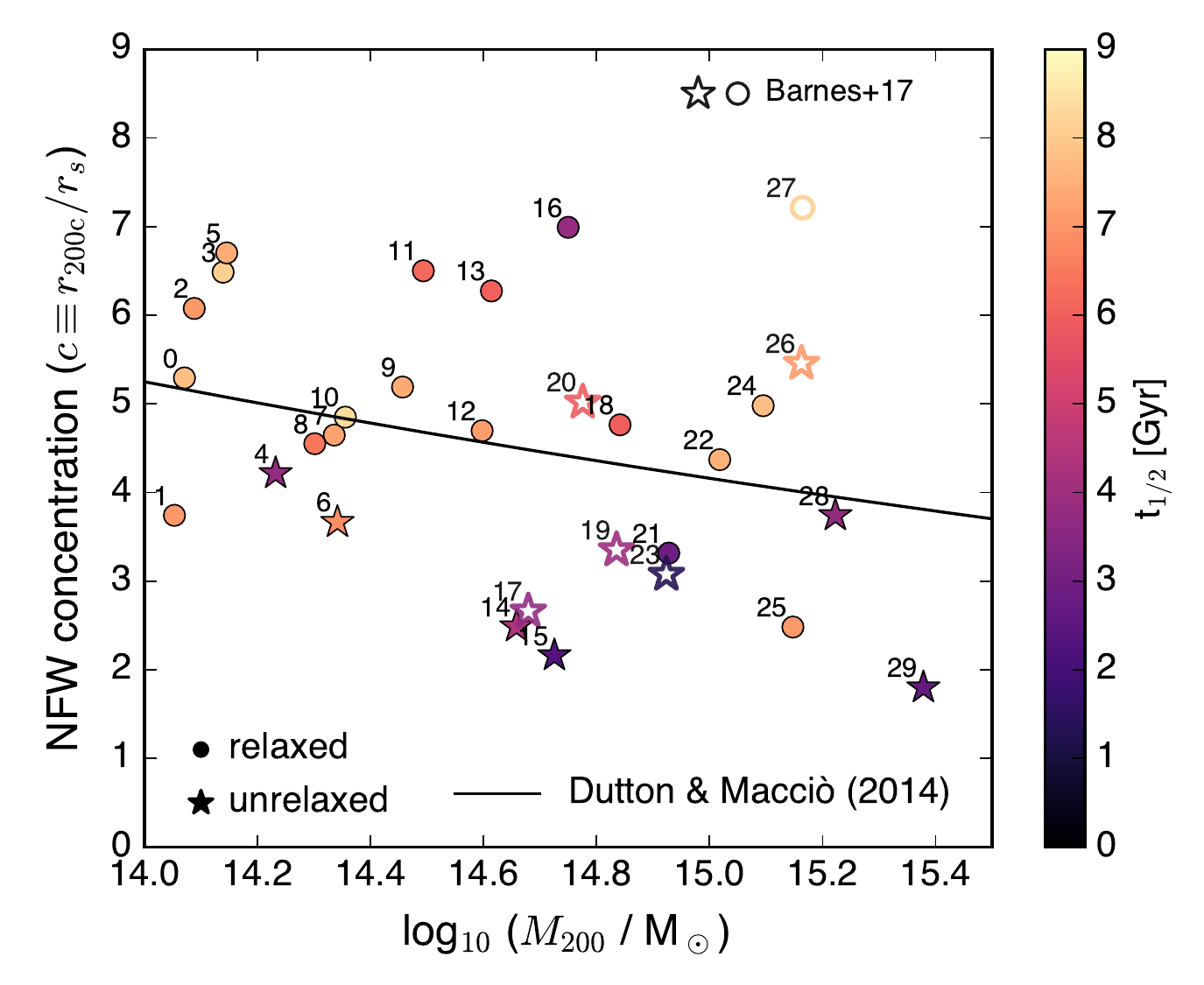}    
       \caption{Mass--concentration relation of the C-EAGLE clusters at redshift $z=0$. The 24 Hydrangea clusters are shown as filled symbols, colour indicating the lookback time when the cluster assembled half its present-day mass, $t_{1/2}$). The additional six clusters introduced by Barnes et al. (in prep.) are shown as empty symbols. Concentration is defined as $c \equiv \rvir/r_s$, where $r_s$ is the best-fit NFW scale radius. Relaxed clusters are shown with circles, unrelaxed ones with star symbols (see text for details). The sample spans a wide range in mass, concentration, dynamical state, and assembly histories.}
      \label{fig:mcplot}
\end{figure}

Fig.~\ref{fig:mcplot} presents an overview of the distribution of the central C-EAGLE clusters in mass--concentration space, where concentrations $c \equiv \rvir/r_s$ were obtained by fitting an NFW profile with scale radius $r_s$ to the spherically averaged dark matter distribution between $r = 0.05\,\rvir$ and $r = \rvir$ \citep{Neto_et_al_2007, Schaller_et_al_2015b}, centred on the potential minimum of the cluster. Clusters that are `relaxed' (i.e.~with an offset between the centre of mass and centre of potential, $s$, less than $0.07\,\rvir$ and a substructure fraction of less than 0.1; \citealt{Neto_et_al_2007}) are shown as circles, unrelaxed haloes that violate one or both of these criteria as stars. Clusters from the Hydrangea sample (i.e.~those with high-resolution regions extending to 10 $\rvir$) are represented by filled symbols, the six remaining C-EAGLE clusters by open symbols. In qualitative agreement with the findings of e.g.~\mbox{\citet{Neto_et_al_2007}}, unrelaxed clusters are typically less concentrated than similarly massive relaxed ones. With significant scatter, the C-EAGLE clusters follow the well-known trend towards lower concentration at higher mass, consistent with the trend from the large DM-only simulation in the Planck $\Lambda$CDM cosmology of \citet{Dutton_Maccio_2014}.

We also indicate the formation time of each cluster, defined as the lookback time when the main progenitor of the cluster assembled half its present-day mass, as the colour of each point. As expected, there is a strong correlation between age and mass in the sense that less massive clusters assembled earlier. A second, albeit less strong, correlation exists between concentration and formation time (less concentrated clusters having typically formed somewhat more recently). In future work, we will exploit this diversity of our cluster sample to investigate in detail the impact of these differences on the galaxy population. Table \ref{tab:info} in Appendix \ref{app:summary} lists the best-fit concentrations along with other information on the masses, positions, and environment of all the C-EAGLE clusters.

In combination, the 24 Hydrangea regions contain, at $z=0$ and within $10\,\rvir$ from the centre of their main halo, 24,442 galaxies with $\mstar \geq 10^9\, \msun$, and 7,207 with $\mstar \geq 10^{10}\, \msun$. We note that this exceeds the corresponding numbers in the 100 cMpc \eagle{} reference simulation by a factor of $\gtrsim$ 2.5.


\section{Stellar masses and quenched fractions of simulated cluster galaxies}
\label{sec:obscomp}

We begin our analysis of the Hydrangea simulations by comparing their predictions for two fundamental galaxy properties to observations, namely their stellar masses (\S \ref{sec:stellarmass}), and quenched fractions (\S \ref{sec:sfr}). We restrict ourselves to comparisons to observations at $z \approx 0$, and will test the simulation predictions at higher redshift in future work. Because the observational studies we are comparing to are focused on the central cluster regions, we include in this section also the six additional C-EAGLE clusters from Barnes et al. (in prep.) whose high-resolution regions extend only to $5\,\rvir$.

\subsection{Galaxy stellar masses}
\label{sec:stellarmass}

The stellar mass of a galaxy is one of its most fundamental characteristics, and many other properties have been shown to correlate strongly with stellar mass: e.g. colour, star formation rate (e.g.~\citealt{Kauffmann_et_al_2003b, Wetzel_et_al_2012}), metallicity (e.g.~ \citealt{Tremonti_et_al_2004, Gallazzi_et_al_2005, Sanchez_et_al_2013}), and, for centrals, their halo mass (e.g.~\citealt{White_Rees_1978}). We now test the galaxy masses predicted by our simulations against observations, for both central cluster galaxies (`BCGs') and their satellites.

\subsubsection{BCG and halo stellar masses}

\begin{figure*}
  \centering
    \includegraphics[width=2.1\columnwidth]{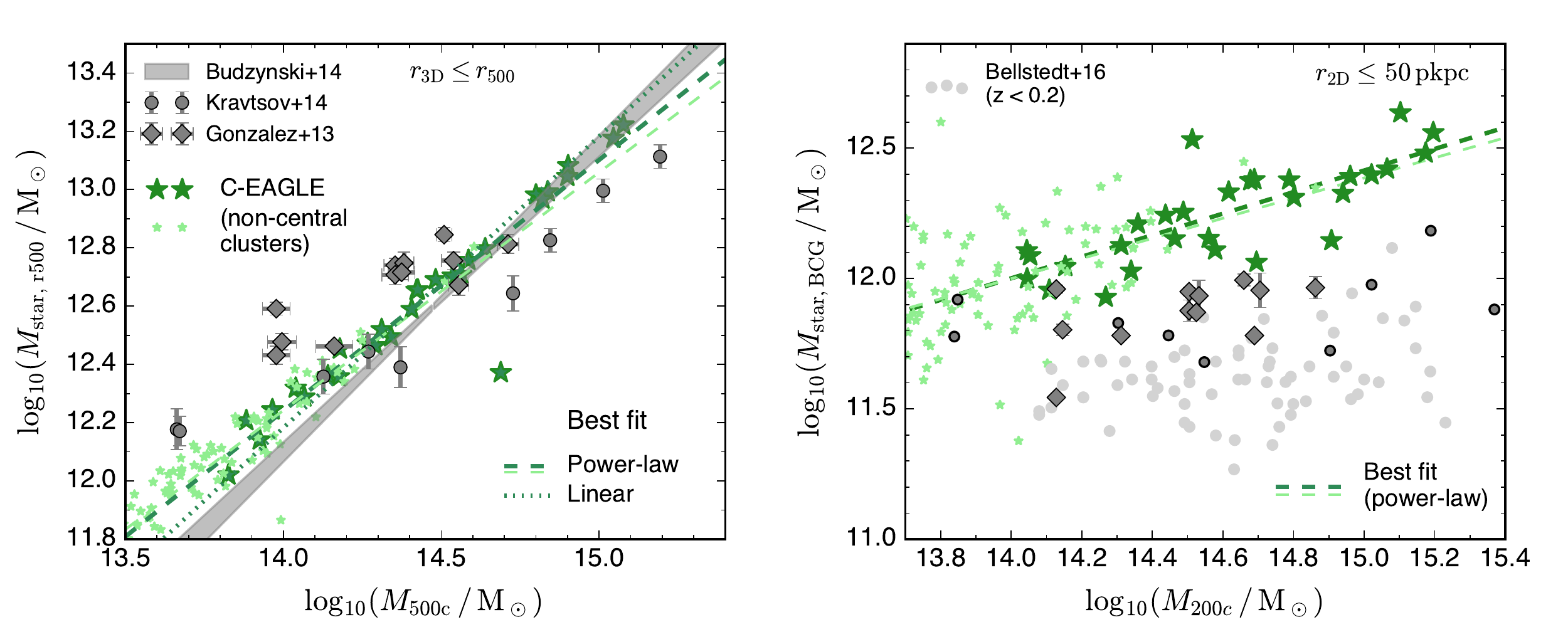}
       \caption{\textbf{Left:} stellar mass of C-EAGLE clusters within $r_{500c}$ (green stars) as a function of true halo mass, compared to several observational data sets (grey points and band). Large dark green symbols represent the 30 central clusters within each simulation, other clusters within the simulation volume (with $\mvir \geq 5\times 10^{13} \msun$) are shown as small light green stars. \textbf{Right:} stellar mass of the simulated BCGs as a function of halo mass, measured within a circular aperture of $R_\text{2D} < 50$ pkpc, compared to observations. In both panels, dashed dark green lines show the best power-law fit to the simulated relation for central clusters with slopes of $\alpha = 0.86 \pm 0.05$ (within $r_{500c}$) and $\alpha = 0.41 \pm 0.06$ (for the BCGs); thin light green lines show the analogous fits for non-central clusters. In the left-hand panel, the dotted dark green line additionally shows the best linear fit, corresponding to a stellar fraction of 1.51 per cent. Although the total mass of stars in the halo (within $r_{500c}$) is reproduced well by the simulations, BCGs are too massive by a factor of $\sim$3. Non-central (`secondary') clusters follow the same relation as their central counterparts.} 
    \label{fig:mstar_r500}
\end{figure*}

In Fig.~\ref{fig:mstar_r500}, we show both the total stellar mass of the clusters in our simulations (i.e.~the mass of all star particles within $r_\text{3D} = r_{500c}$, the radius within which the average density equals 500 times the critical density; left-hand panel), and the stellar mass of the BCG, i.e.~the galaxy at the potential minimum of the cluster's FoF halo, in the right-hand panel (integrated within a circular aperture with $R_\mathrm{2D} = 50$ pkpc, see below). Predictions from our simulations are shown in shades of green, dark for the 30 central clusters (i.e.~the most massive ones in their simulation volume), and light green for others. Observational data are show in grey. For halo stellar masses, we compare to the observations of \citet{Gonzalez_et_al_2013} and \citet{Kravtsov_et_al_2014}, and the best-fit relation derived from SDSS images by \citet{Budzynski_et_al_2014}. In the observations, $M_\mathrm{500c}$ is estimated from the X-ray temperature \citep{Gonzalez_et_al_2013, Kravtsov_et_al_2014} and the mass-richness relation \citep{Budzynski_et_al_2014}; we multiply these with a factor of 1.5 to convert from $M_{500c}$ to $\mvir$. In the simulations, we measure halo masses directly (masses derived from mock X-ray spectra are presented in Barnes et al., in prep.). Note that the first two observational datasets are from clusters at $z \lesssim 0.1$, whereas the \citet{Budzynski_et_al_2014} relation was derived for clusters at $0.15 \leq z \leq 0.4$. We here compare to the simulation output at $z=0.101$ as a compromise between these two ranges.  

We first consider the simulation prediction for the 30 central clusters (dark green stars in Fig.~\ref{fig:mstar_r500}), which exhibit a fairly tight relation between halo mass and both the halo and BCG stellar mass. The former is slightly sub-linear (best-fit power law index $\alpha = 0.86 \pm 0.05$, with a best-fit overall stellar fraction of 1.51 per cent). This is less steep than the relation of \citet{Budzynski_et_al_2014}, $\alpha = 1.05 \pm 0.05$, but slightly steeper than in \citet{Gonzalez_et_al_2013} and \citet{Kravtsov_et_al_2014}. We therefore conclude that, overall, the (central) C-EAGLE clusters have formed approximately realistic amounts of stellar mass (see also Barnes et al., in prep.).

The agreement is less good when only the stellar mass of the BCG is considered, which we define as the mass within a (2D) radial aperture of 50 pkpc and integrating through the entire high-resolution simulation region along the line of sight (right-hand panel of Fig.~\ref{fig:mstar_r500}). \citet{Stott_et_al_2010} have shown that this aperture mimicks the \citet{Kron_1980} aperture commonly encountered in observational analyses, including the one of \citet{Bellstedt_et_al_2016} whose BCG stellar mass measurements (at $z < 0.2$) we show as light grey circles. Also shown are BCG masses from \citet{Kravtsov_et_al_2014}, measured within a projected radius of 50 pkpc, and those of \citet{Gonzalez_et_al_2013}, corrected to $R_\text{2D} \leq 50$ pkpc by multiplying with a correction factor of 0.4 (see \citealt{Gonzalez_et_al_2005}). 

The stellar masses of the simulated central BCGs (dark green) lie significantly above all these datasets, by $\sim$0.3 dex compared to \citet{Gonzalez_et_al_2013} and \citet{Kravtsov_et_al_2014}, and $\sim$0.6 dex compared to \citet{Bellstedt_et_al_2016}. This discrepancy is greatest for the most massive haloes. In the companion paper of Barnes et al. (in prep.), we demonstrate that our clusters also have a hot gas fraction that is somewhat too high compared to what is inferred from X-ray observations. Correspondingly, the star formation rates of the central cluster galaxies within the central 15 pkpc (not shown) are all in the range from $\sim$1 to $\sim$10 $\msun\,\text{yr}^{-1}$, whereas only $\lesssim$ 50 per cent of observed central cluster galaxies show evidence for star formation at this level (e.g.~\citealt{Hoffer_et_al_2012, Donahue_et_al_2015, Fogarty_et_al_2015}). It is tempting to identify this excess star formation as the cause of the unrealistically high BCG masses. However, only $\sim$10 per cent of the mass of our simulated BCGs has typically been formed at $z < 1$. The BCG masses are therefore not predominantly too high because of artificially high levels of in-situ star formation at low redshift, but reflect a shortcoming of the simulations in modelling their high-redshift proto-cluster progenitors. 

Due to their large volume, the Hydrangea simulations also contain a large number of `secondary' cluster haloes that are less massive than the `primary' one at the centre of each simulation. In total, there are 38 of these with $\mvir > 5\times 10^{13} \msun$ within 10 (5) $\rvir$ from the central cluster in the Hydrangea (other C-EAGLE) simulations. This number is boosted to 81 when including objects beyond this nominal edge of the high-resolution sphere, but which are still far away ($>$ 8 pMpc) from any low-resolution boundary particles\footnote{These `external' secondary clusters can exist because the high-resolution regions at $z \approx 0$ are, in general, non-spherical. We have verified that they do not display any significant difference in their stellar masses from secondary clusters within the nominal high-resolution region.}. At fixed $M_\mathrm{500c}$, secondary clusters contain the same stellar mass as primaries, both within $r_{500c}$ and in their BCG.

\subsubsection{The stellar mass function of satellite galaxies}
We now compare the simulation predictions for the low-redshift satellite galaxy stellar mass function (GSMF). This has been studied observationally by several authors in recent years, including \citet[based on SDSS spectroscopic data and the \makebox{\citealt{Yang_et_al_2007}} SDSS halo catalogue]{Yang_et_al_2009}, \citet[from the WINGS survey of nearby galaxy clusters]{Vulcani_et_al_2011}, and \citet[again from SDSS data but stacking galaxy counts around bright isolated galaxies]{Wang_White_2012}.

All three of these observational studies exclude BCGs, but each uses a somewhat different definition of `satellite galaxy'. We therefore begin by briefly describing these different selections and our methods for approximating them within the C-EAGLE simulations.

\citet{Yang_et_al_2009} used the \citet{Yang_et_al_2007} halo catalogue to match SDSS galaxies to underlying dark matter haloes based on their spatial distribution. The most massive galaxy in each halo is identified as `central', while all others are `satellites'. These authors report the satellite GSMF for different bins of halo mass, out of which we here compare to the (most massive) bin with $14.4 \leq \log_{10} M_{200}/ (h^{-1}\,\msun) < 14.7$. There are 7 C-EAGLE clusters in this mass range, for which we select all simulated galaxies that \textsc{subfind} identifies as satellites of the cluster FoF halo.

\citet{Vulcani_et_al_2011} assigned cluster membership in the WINGS catalogue \citep{Fasano_et_al_2006} based on 2D projected distance from the cluster centre ($R_{2D} \leq 0.6 r_{200}$). The WINGS clusters have $\mvir \gtrsim 10^{14.5}\, \msun$ \citep{Fasano_et_al_2006}\footnote{We have used the $L_X$--$M_\mathrm{500c}$ relation of \citet{Vikhlinin_et_al_2009a} to convert the WINGS X-ray luminosities of \citet{Fasano_et_al_2006} to halo masses, with an additional correction factor of 1.5 to convert to $\mvir$.}. We therefore compare to the 17 C-EAGLE clusters with $\mvir \geq 10^{14.5}\, \msun$ and select those galaxies within $R_{2D} \leq 0.6 \rvir$ from the potential minimum of each cluster\footnote{We have not imposed an additional cut along the line-of-sight, because the criterion of $\Delta z \leq 3 \sigma$ (with redshift $z$ and cluster velocity dispersion $\sigma$) of \citet{Vulcani_et_al_2011} corresponds to an integration length that is comparable to the size of the high-resolution region in our simulations.}. 

\citet{Wang_White_2012} used a fixed 300 pkpc aperture around bright isolated galaxies to count satellites, but even in their highest stellar mass bin, the typical halo mass (as estimated from semi-analytic models) is only $\sim$$10^{13.7}\, \msun$. This is slightly lower than the halo mass range of our simulations, so we compare to simulated haloes in the mass range $14.0 \leq \log_{10} \mvir/\msun < 14.5$ (13 clusters) and re-normalize the \citet{Wang_White_2012} GSMF as described below.

Besides differences in galaxy selection, the observations span a range of redshifts, with median $z \approx 0.1$ for SDSS \citep{Yang_et_al_2009, Wang_White_2012}, while the WINGS clusters lie at $0.04 < z < 0.07$ \citep{Vulcani_et_al_2011}. For simplicity, we compare all three datasets to the simulation predictions at $z=0.101$, but have verified that differences to the predictions at $z=0$ are small. In all three cases, we compute stellar masses in the simulations as the sum of all gravitationally bound star particles that are within 30 pkpc from the potential minimum of their subhalo. \citet{Schaye_et_al_2015} have shown that this aperture yields a good match to the Petrosian apertures often employed in observations, including those from the SDSS. We restrict our comparison here to the primary (central) clusters of each simulation.

\begin{figure}
  \centering
    \includegraphics[width=1.02\columnwidth]{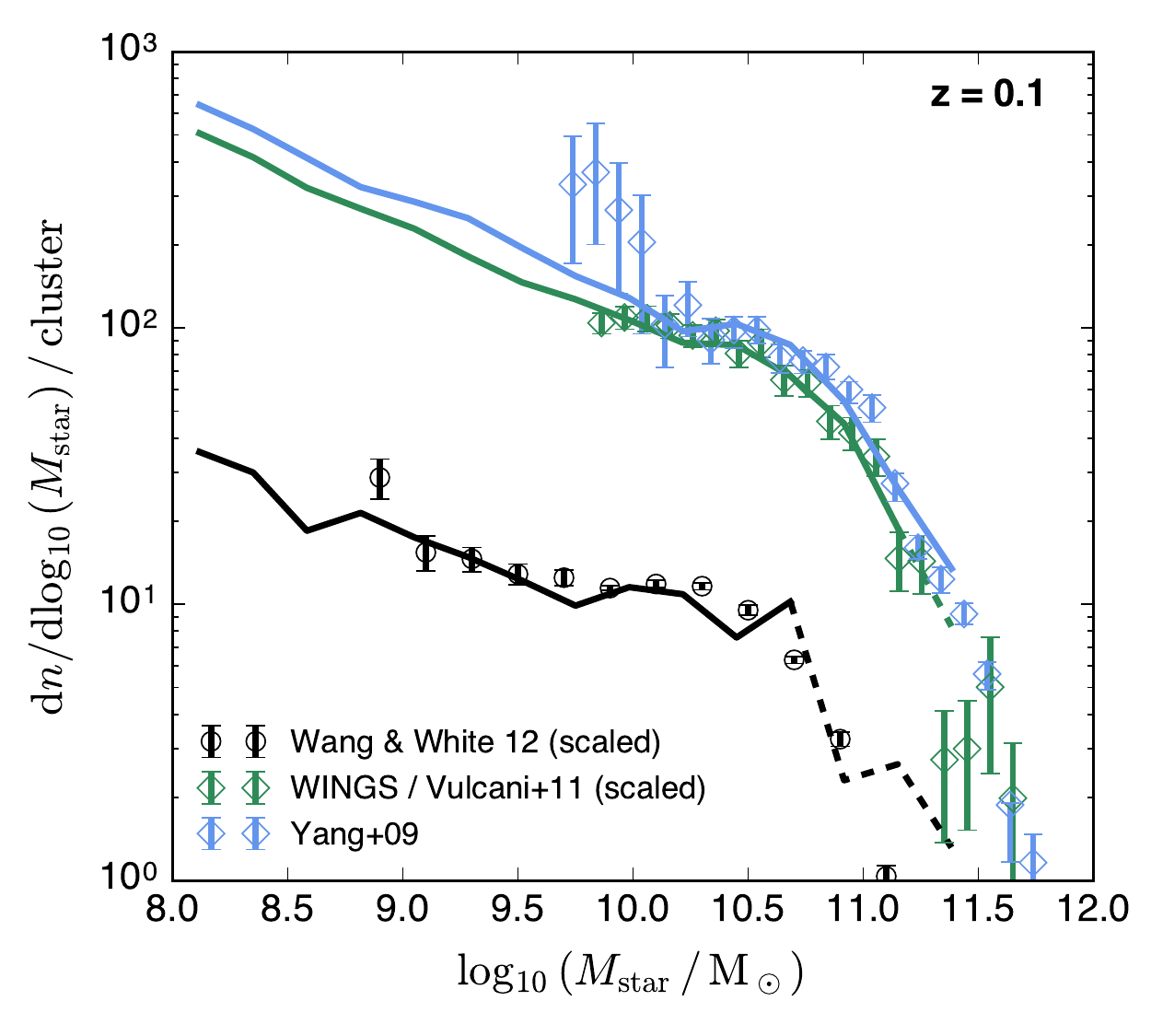}
       \caption{Galaxy stellar mass function (GSMF)  at $z=0.101$ for satellites in the C-EAGLE simulations (solid lines, dashed where there are $< 10$ galaxies per 0.25 dex bin) compared to observations (open diamonds). The three different lines represent galaxy selections approximately matched to the respectively-coloured observational survey: $14.0 \leq \log_{10} (\mvir/\msun) < 14.5$, $R_\text{2D} < 300$ pkpc (black); $14.5 \leq \log_{10} (\mvir/\msun)$, $R_\text{2D} < 0.6 \rvir$ (green); $14.4 \leq \log_{10} \mvir/(\msun h^{-1}) < 14.7$, all halo members (blue). Overall, the simulations achieve an excellent match to the observations.}
    \label{fig:gsmf_obscomp}
\end{figure}

The comparison between simulations and observational data is shown in Fig.~\ref{fig:gsmf_obscomp}. The simulated GSMF is shown with solid lines where bins contain more than ten galaxies, and dashed lines for more sparsely sampled bins at the high stellar mass end. The observations are shown as empty symbols, with error bars indicating the observational $1\sigma$ uncertainties. Data points for WINGS (green) and \citet{Wang_White_2012} have been scaled by multiplying their stellar mass function with a correction factor such that the total number of galaxies above a given threshold ($10^{9.8} \msun$ for WINGS, $10^{9.4} \msun$ for \citealt{Wang_White_2012}) is the same as in the C-EAGLE simulations. In the case of WINGS, this is necessary because the GSMF presented by \citet{Vulcani_et_al_2011} has been scaled for the purpose of comparing to field galaxies, while the GSMF of \citet{Wang_White_2012} was derived for haloes that are less massive than the C-EAGLE sample (see above). Differently coloured lines correspond to simulated GSMFs matched to the correspondingly coloured observational data set, as described above.

Overall, the simulated $z \approx 0$ GSMF agrees well with all three data sets. The only slight tension is seen at the low-mass end of the \citet{Wang_White_2012} and \citet{Yang_et_al_2009} comparisons, where the observations hint at an upturn of the GSMF that is not seen in the simulations. We note that these observational data points also have large uncertainties -- in the case of \mbox{\citet{Yang_et_al_2009}}, the discrepancy for an individual data point is only significant at the $\sim$1$\sigma$ level -- but alternatively, this deficiency might be a consequence of overly efficient star formation quenching in low-mass galaxies in our simulations, as we shall discuss shortly. 

The accuracy of the predicted cluster GSMF reflects, in part, the calibrated match between the \eagle{} simulations and the field GSMF \citep{Schaye_et_al_2015}. However, as shown below, there are significant differences between the field and cluster GSMF in our simulations. The close agreement between our cluster GSMF and the observations shown in Fig.~\ref{fig:gsmf_obscomp} therefore suggests a realistic modelling of cluster-specific aspects of galaxy formation, at least to the extent that they manifest themselves in the stellar mass of galaxies. We exploit this success of our simulations further in \S \ref{sec:gsmf}, where we compare the GSMF in and around simulated clusters to the field.

\subsection{Satellite quenched fractions}
\label{sec:sfr}

A second key property of galaxies, which is closely related to their stellar mass, is their star formation rate. Observations have shown conclusively that galaxies in dense environments are biased towards lower specific star formation rates (sSFR $\equiv$ SFR/$\mstar$;  e.g.~\citealt{Kauffmann_et_al_2004}), with the key difference being an increased fraction of passive galaxies (e.g.~\citealt{Peng_et_al_2010, Wetzel_et_al_2012}). We now test the C-EAGLE predictions for the quenched fraction of simulated satellites.

\label{sec:fquench_sat}

\begin{figure*}
  \centering
    \includegraphics[width=2.1\columnwidth]{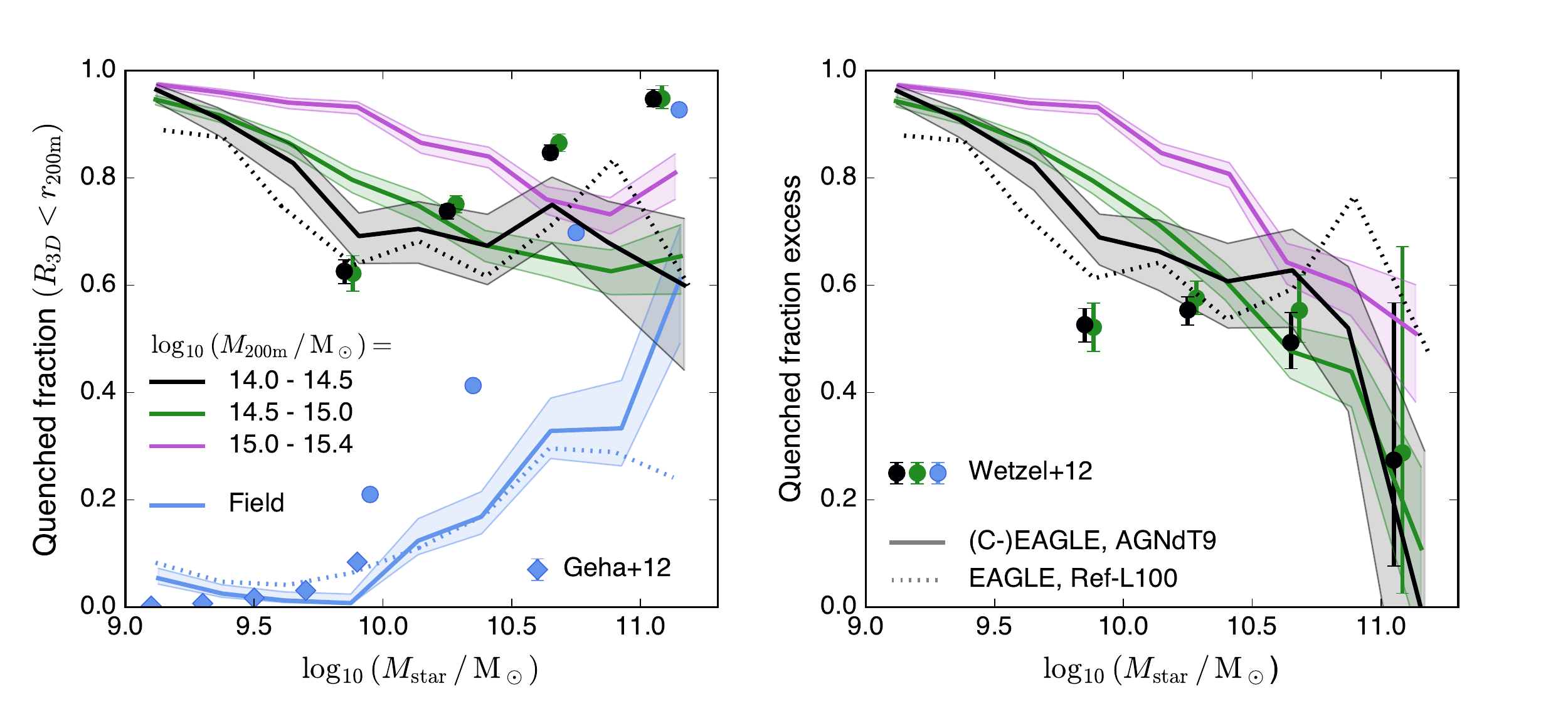}
       \caption{\textbf{Left:} Quenched satellite fraction within $r_\text{3D} \leq r_{200m}$, in bins of cluster mass (differently coloured solid lines) as a function of stellar mass. The blue solid line shows the corresponding trend in the field, i.e.~centrals in the AGNdT9-L050 simulation from the \eagle{} suite. Shaded bands indicate $1\sigma$ binomial uncertainties \citep{Cameron_2011}. The dotted blue and black lines are the corresponding trends from the \eagle{} Ref-L100 simulation. Filled circles with error bars show the corresponding values from the SDSS DR7 analysis of \citet{Wetzel_et_al_2012} and blue diamonds the observed quenched fractions of field dwarfs from \citet{Geha_et_al_2012}. In agreement with observations, simulated satellites show an enhanced quenched fraction compared to the field, albeit with discrepancies in the trends with $\mstar$ (see text for details). \textbf{Right:} The satellite quenched fraction excess, $(f_q^\text{sat} - f_q^\text{cen}) / (1-f_q^\text{cen})$, which shows quantitative agreement between simulations and observations at $\mstar > 10^{10} \msun$.} 
           \label{fig:fquenched_obscomp}
\end{figure*}

In the left-hand panel of Fig.~\ref{fig:fquenched_obscomp}, we show the passive fraction of C-EAGLE cluster satellites as a function of stellar mass and host mass. For consistency with the observational analysis of \mbox{\citet{Wetzel_et_al_2012}}, we define `passive' galaxies as those with sSFR $< 10^{-11} \text{yr}^{-1}$. For the same reason, we take cluster mass here as $M_{200m}$ (the mass within the radius $r_{200m}$ inside which the average density is 200 times the \emph{mean}, as opposed to critical, density of the Universe) and select as satellites those galaxies at radii $r_{3D} \leq r_{200m}$ (excluding the BCG)\footnote{The group finding algorithm of \citet{Wetzel_et_al_2012} accounts for line-of-sight projection in a probabilistic way, with the aim of assigning galaxies to haloes in 3D space. We have repeated the analysis presented here with a cut in $R_\text{2D}$ instead, and found no qualitative differences.}. 

Clusters are grouped into three mass bins between $M_\mathrm{200m} = 10^{14}\, \msun$ and $10^{15.5}\, \msun$, represented by different colours. For comparison, we also show the corresponding quenched fraction of central galaxies from the \eagle{} AGNdT9-L050 simulation (which was run with the exact same simulation parameters as C-EAGLE; solid blue line). Shaded bands indicate the statistical binomial $1\sigma$ uncertainty \citep{Cameron_2011} on the quenched fraction. Observational data from \citet{Wetzel_et_al_2012} are overlaid as filled circles in corresponding colours; the error bars represent $1\sigma$ uncertainties. We note that their observations do not probe the highest halo mass bin (purple). Also plotted are the quenched fractions of low-mass field galaxies from \citet[blue diamonds]{Geha_et_al_2012}. Finally, the analogous trends from the \eagle{} Ref-L100 simulation -- whose parameters describing AGN feedback are different from C-EAGLE, see Section \ref{sec:eagle} -- are shown as dotted lines, both for centrals (blue) and the lowest-mass cluster bin (black).

The dominant feature of Fig.~\ref{fig:fquenched_obscomp} is an increased quenched fraction of satellites across the range of halo masses shown here ($M_{200m} > 10^{14} \msun$), at least at $\mstar < 10^{11} \msun$, which agrees qualitatively with observations. Similar to what is seen in the \citet{Wetzel_et_al_2012} data, the quenched fractions in the 14.0--14.5 and 14.5--15.0 halo mass bins (black/green) closely follow each other. For the 9 clusters with $M_\text{200m} > 10^{15}\, \msun$, the simulations predict a substantially higher quenched fraction, especially at intermediate stellar masses ($\mstar \approx 10^{10} \msun$). 

At $\mstar \gtrsim 10^{10} \msun$, the quenched satellite fraction in the C-EAGLE simulations is lower than observed. This discrepancy is most severe for the most massive galaxies ($\mstar \approx 10^{11.5}\, \msun$; 70 per cent in C-EAGLE vs.~near unity in the data). We point out, however, that the same is true for \emph{central} galaxies, in both the \eagle{} AGNdT9 and Ref runs (blue solid and dotted lines), which points to a more fundamental discrepancy between the simulations and observations, either because quenching due to internal mechanisms such as AGN (see e.g.~\citealt{Bower_et_al_2017}) is not efficient enough in the \eagle{} model, or because the quenched fractions in the observations are overestimated (as demonstrated by Trayford et al., in prep., in the case of quenched fractions derived from galaxy colours)\footnote{We note, however, that the quenched fractions of \citet{Wetzel_et_al_2012} are derived from optical spectra and not colours. A recent study by \citet{Chang_et_al_2015} found that these tend to \emph{over}estimate SFRs, which would exacerbate rather than alleviate the discrepancy.}. To isolate the \emph{environmental} impact on the quenched fraction, we plot in the right-hand panel of Fig.~\ref{fig:fquenched_obscomp} the `quenched fraction excess', defined as $(f_q^\text{sat} - f_q^\text{cen}) / (1-f_q^\text{cen})$ as proposed by \citet{Wetzel_et_al_2012}. In this metric, the simulations show much closer agreement with the observations, indicating that the environmental impact on star-forming gas is modelled correctly in our simulations, at least for $\mstar > 10^{10} \msun$.

At lower stellar masses ($\mstar \lesssim 10^{10} \msun$), observations indicate a continued decrease in the passive fraction of both satellites and centrals with decreasing stellar mass \citep[blue diamonds in Fig.~\ref{fig:fquenched_obscomp}]{Geha_et_al_2012}. While this is approximately reproduced by \eagle{} centrals -- whose passive fraction is $< 10$ per cent at $\mstar = 10^9\, \msun$ -- the passive fraction of satellites in our simulations \emph{increases} significantly, and almost reaches unity at $\mstar = 10^9\, \msun$, independent of host mass. In \citet{Schaye_et_al_2015}, it was already shown that EAGLE predicts a passive fraction in the combined galaxy population (centrals and satellites) that rises towards lower stellar masses below $\mstar \approx 10^{9.5} \msun$, and that this effect is strongly resolution dependent. Because almost all these quenched low-mass galaxies are satellites (at least down to $\mstar = 10^9 \msun$, see our Fig.~\ref{fig:fquenched_obscomp}), the over-efficient quenching of low-mass satellites in C-EAGLE can therefore also be primarily ascribed to resolution effects, even though all galaxies shown here are resolved by $\gg 1000$ particles.

We speculate that this effect may be connected to the overly porous structure of atomic hydrogen discs in many EAGLE galaxies reported by \citet{Bahe_et_al_2016}. As a consequence of limited resolution, star formation feedback events in the EAGLE model create holes that are larger than observed, and it is possible that this increased porosity might make the disc more susceptible to being stripped under the influence of ram pressure.


\section{Environmental influence on stellar masses}
\label{sec:gsmf} 

We have shown in the previous section that the C-EAGLE simulations produce realistic satellite galaxy stellar mass distributions in the cores of massive clusters, while the underlying \eagle{} model reproduces, by construction, the GSMF in the field \citep{Schaye_et_al_2015, Crain_et_al_2015}. This gives the simulations power to gain theoretical insight into how environment affects the GSMF in and around clusters. We will now proceed with a first analysis of these environmental effects. For consistency with the previous section and SDSS-based observations (see \citealt{Schaye_et_al_2015}), we continue to only consider stars within 30 pkpc from the potential minimum of each galaxy's subhalo. Despite some difference in detail, none of our findings below change qualitatively when including all stars bound to the subhalo instead.

\subsection{Environmental impact on the normalized stellar mass function}
A key difficulty of comparing galaxy stellar mass functions between different environments is the application of a suitable normalization, since by definition the overall density of galaxies is higher in clusters than in the field. In the observational literature this has, for instance, been accomplished by re-normalizing cluster and field mass functions so both yield the same total number of galaxies above a given mass limit (e.g. \citealt{Vulcani_et_al_2011}). In our simulations, a more natural way to accomplish this is to divide the number of galaxies by the \emph{total mass} within the same volume, effectively computing the bias of galaxies of different mass with respect to the general mass distribution as in \citet{Crain_et_al_2009}.

\begin{figure*}
  \centering
    \includegraphics[width=2.1\columnwidth]{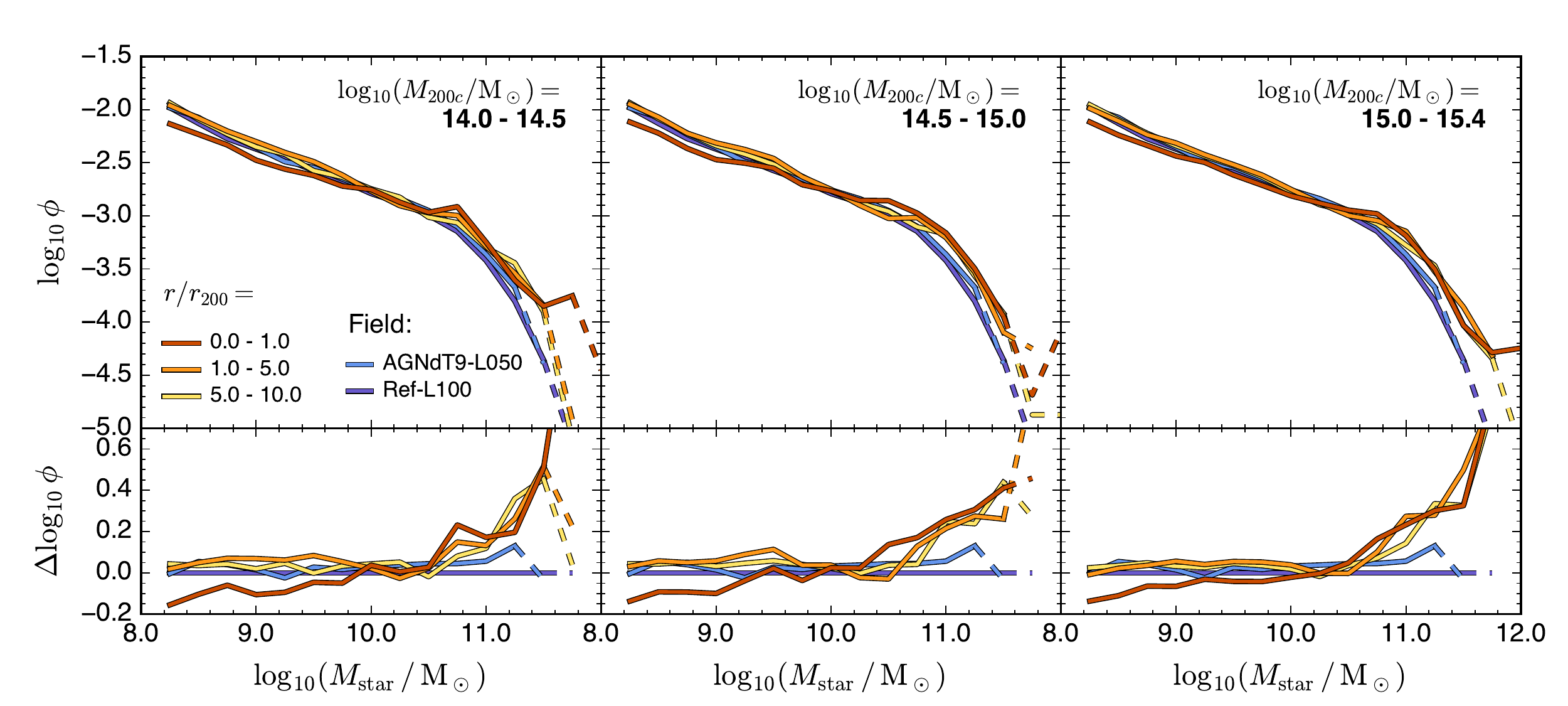}
       \caption{\textbf{Top row:} galaxy stellar mass function (GSMF) normalised to the total mass within the respective volume, $\phi \equiv \mathrm{d}n/\mathrm{d}\log_{10}M_\text{star} / (M_\text{tot}/10^{10} \mathrm{M}_\odot)$. Individual columns contain clusters of different $\mvir$ (as indicated in the top-right corner). Differently coloured lines (dashed where there are less than ten galaxies per 0.25 dex bin) represent different radial zones in each cluster: inside $\rvir$ (red); between 1 and 5 $\rvir$ (i.e.~the region containing a population of backsplash galaxies, orange); the far outskirts beyond 5 $\rvir$ (yellow). For comparison, the mass functions from the AGNdT9-L050 (blue) and Ref-L100 (purple) \eagle{} runs are also shown. \textbf{Bottom row:} logarithmic ratio between each GSMF and that from the Ref-L100 periodic box. All halo mass bins show an excess of massive galaxies in and around clusters, without a clear radial trend. Galaxies less massive than $\sim$$10^{10}\, \msun$, on the other hand, are deficient in the central cluster regions (red).}
    \label{fig:gsmf_bias}
\end{figure*}

We present this comparison in Fig.~\ref{fig:gsmf_bias}, where we show the normalized GSMF, i.e.~$\phi \equiv \text{d}n/\text{d}\log_{10} \mstar\,/\,(M_\text{tot}/10^{10}\,\msun)$, where $M_\text{tot}$ is the total mass within the volume that galaxies are selected from. We distinguish between Hydrangea clusters in three different mass bins (different panels, increasing from left to right). For each bin, we stack all clusters and extract the GSMF in three concentric shells centred on the cluster's potential minimum, in the radial range $r = $ 0--$\rvir$ (the virialized central region), 1--5 $\rvir$ (the region comprising a mix of first-infall and backsplash galaxies; see \citealt{Bahe_et_al_2013}), and 5--10 $\rvir$ (the primordial infall region). For comparison, we also show the normalized GSMF from the \eagle{} AGNdT9-L050 (blue) and Ref-L100 (purple) periodic box simulations. Since these model representative cosmic volumes, they can be taken as estimates of the `field' GSMF. There is very close agreement between these latter two distributions \makebox{\citep{Schaye_et_al_2015}}, with the key difference being that Ref-L100 extends to higher masses due to its eight times larger volume.

At first sight, the normalized stellar mass function shows little difference between the different environments, with particularly close agreement at $\mstar \approx 10^{10}\, \msun$ in all three halo mass bins. On closer inspection, however, there are two clear and significant differences. Firstly, there is a deficiency of low stellar mass galaxies ($\mstar \lesssim 10^{10}\, \msun$) within $\rvir$ (red line), of up to $\sim$0.2 dex. Secondly, massive galaxies ($\mstar \gtrsim 10^{10.5}\, \msun$) are more numerous in our simulated clusters, from the central region ($< \rvir$) to the far outskirts (the 5--10 $\rvir$ zone; yellow). Qualitatively, this is consistent with the recent Dark Energy Survey analysis of \citet{Etherington_et_al_2017}, who found a higher fraction of massive galaxies in higher-density environments. The bottom panels show the mass functions normalized to Ref-L100 to bring out these differences more clearly. Over more than a decade in halo mass, the environmental differences in galaxy stellar mass show no strong dependence on cluster mass.

\begin{figure*}
  \centering
    \includegraphics[width=2.1\columnwidth]{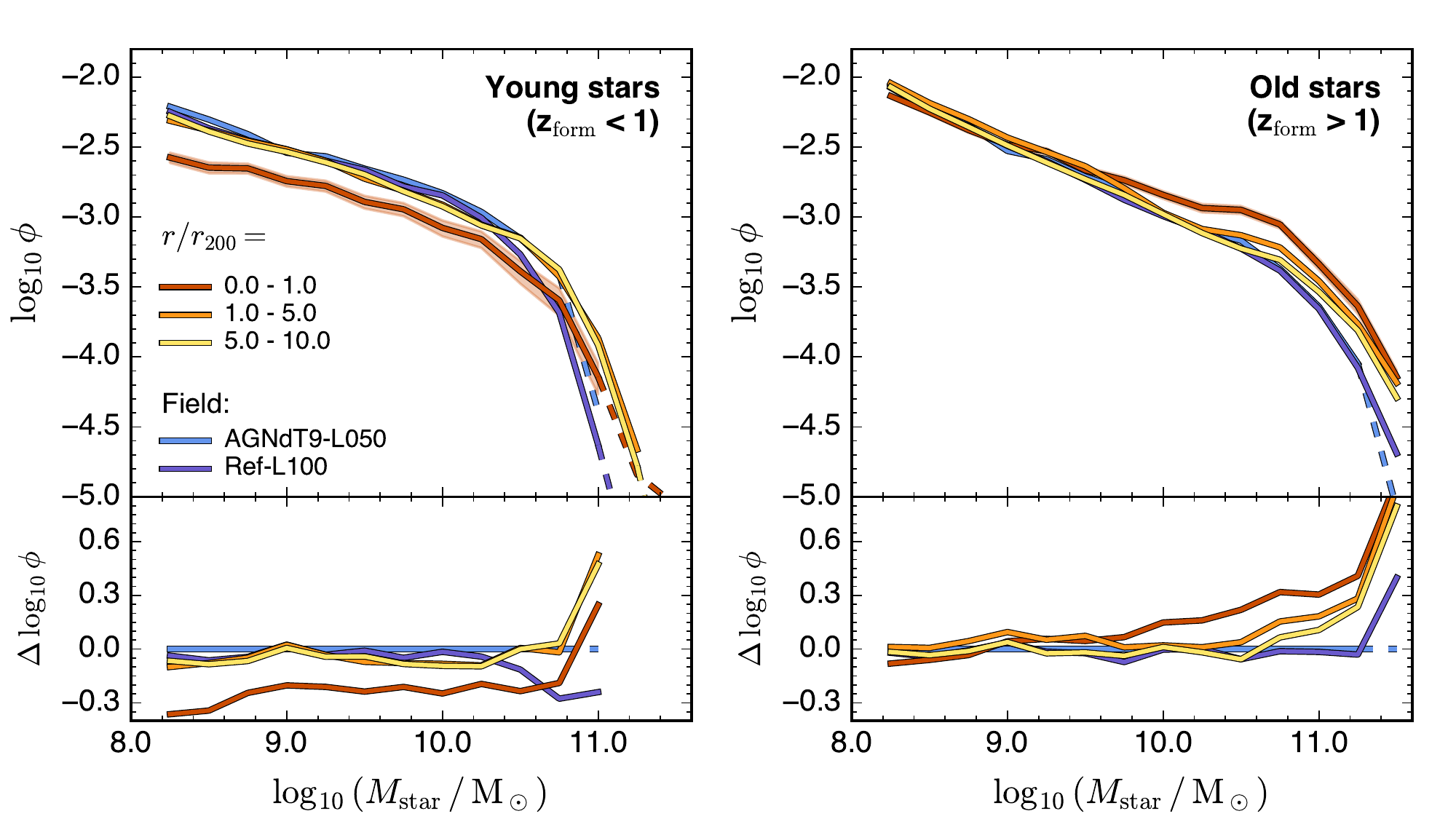}
       \caption{\textbf{Top panels:} the normalized $z=0$ galaxy stellar mass functions $\phi \equiv \mathrm{d}n/\mathrm{d}\log_{10}M_\text{star} / (M_\text{tot}/10^{10} \mathrm{M}_\odot)$ split by stellar formation redshift into `young' stars ($z_\text{form} < 1.0$, left-hand panel) and `old' stars ($z_\text{form} > 1.0$, right-hand panel). As in Fig.~\ref{fig:gsmf_bias}, the Hydrangea volumes are split into three radial zones (red, orange, and yellow lines) and compared to the \eagle{} periodic-box simulations (blue/purple); dashed lines indicate bins with fewer than ten galaxies. The shaded band, shown only for $r < \rvir$ (red) for clarity, indicates the Poisson uncertainty on the GSMF. For clarity, the \textbf{bottom panels} show the logarithmic ratio between each GSMF and that from the AGNdT9-L050 periodic box. Evidently, the deficiency at low mass and excess of stellar mass at high mass are due to two different processes, since the former only affects young, and the latter mostly old stars.}
    \label{fig:gsmf_bias_z1}
\end{figure*}

The deficiency of low-mass galaxies within the virial radius can be due to tidal stripping (or even complete disruption) of satellites, lack of stellar mass growth as a result of star formation quenching, or a combination thereof. In Fig.~\ref{fig:gsmf_bias_z1} we test these hypotheses by constructing galaxy stellar mass functions separately for young and old stars, defined as those formed after or before redshift $z=1$, respectively. The environmental impact on these two different populations is strikingly different. From the left-hand panel, the young stellar mass function shows a strong deficiency at the low-mass end (by up to 0.4 dex), but only a minor high-mass excess except for the most massive galaxies ($M_\text{star, young} > 10^{11}\,\msun$). From the horizontal offset between the curves, stellar stripping would have to reduce the young stellar mass of an $M_\text{star, young} = 10^{10}\,\msun$ galaxy by $\sim$0.3 dex to account for this offset. However, we will show in a forthcoming paper that stellar stripping within $\rvir$ has a typical effect of $< 0.1$ dex at these mass scales and can therefore not be a significant contributor to the lack of young stars within $\rvir$. 

In contrast, galaxies with a given mass in old stars down to $\sim$$10^8\, \msun$ are equally common within the virial radius as in the field (right-hand panel); this suggests that complete disruption of galaxies within $\rvir$ is, likewise, not a significant contributor to the deficiency of low-mass galaxies in clusters. We therefore conclude that this is \emph{predominantly due to the effect of star formation quenching}, which reduces the late-time growth of galaxies within $\rvir$.   

At the high-mass end, the right-hand panel of Fig.~\ref{fig:gsmf_bias_z1} demonstrates that the excess of galaxies is largely due to old stars. Their excess shows a systematic trend with radius, in the sense that galaxies with a high mass in old stars are most highly overabundant within $\rvir$, but a clear effect remains even at $r > 5\rvir$ from the cluster centre. We will return to this in Section~\ref{sec:msub_mstar} below.

\subsection{The galaxy--subhalo connection in and around clusters}
\subsubsection{Subhalo mass functions}

\begin{figure}
  \centering
    \includegraphics[width=\columnwidth]{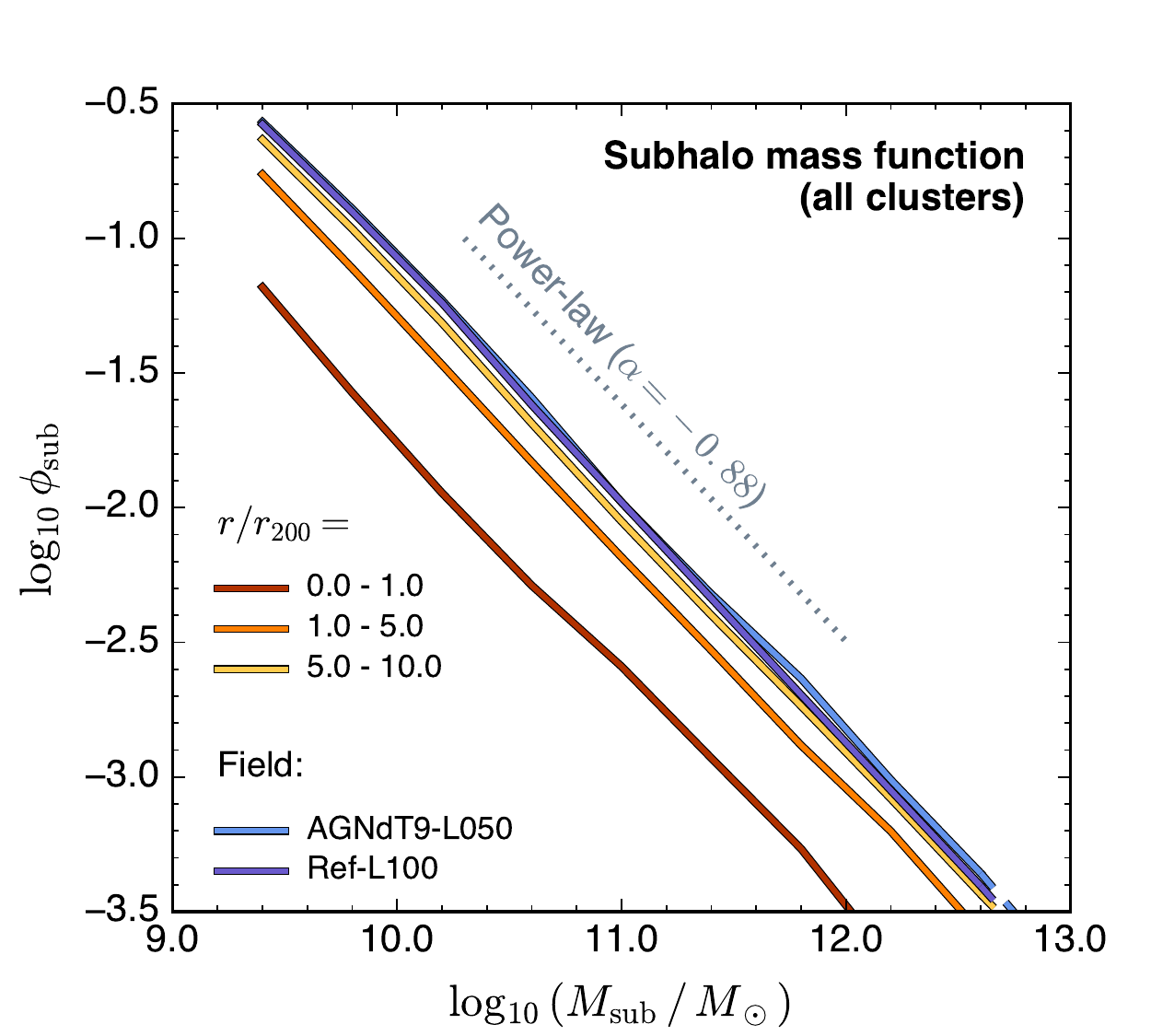}
       \caption{The normalized \emph{subhalo} mass function (including central subhaloes) in and around the Hydrangea clusters, in analogy to the GSMF presented in  Fig.~\ref{fig:gsmf_bias}. Masses are computed as the sum of all particles that are gravitationally bound to a given subhalo (including centrals). Different colours indicate different simulation zones and each mass function is normalized to the total mass in its respective zone. In contrast to the galaxy stellar mass function, all subhalo mass functions are near-perfect power-laws with a slope of approximately -0.88 (grey dotted line).}
    \label{fig:subhalo_bias}
\end{figure}

An excess of massive galaxies in the vicinity of galaxy clusters may not be unexpected in $\Lambda$CDM, because the addition of large- and small-scale density peaks lead to earlier collapse of haloes, i.e.~`assembly bias' (e.g. \citealt{Gao_et_al_2005, Gao_White_2007}). We test the importance of this effect in Fig.~\ref{fig:subhalo_bias}, where we show the \emph{subhalo} mass function, again comparing different zones in our cluster simulations and the periodic box volumes from \eagle{}, normalized by their total mass. Recall from above that our definition of `subhalo' also includes the most massive bound structure within a FOF halo, i.e.~the one hosting the central galaxy.

The subhalo mass functions differ markedly from the galaxy stellar mass function, and follow an almost perfect power-law over 4 orders of magnitude in subhalo mass (from $\sim$$10^9$ to $\sim$$10^{13} \msun$). A power-law subhalo mass function agrees with previous cluster simulation studies, although there is a mild difference between the slopes. Our simulations yield a slope of $\alpha \approx -0.88$ (see also \citealt{Despali_Vegetti_2016} for the subhalo mass function in \eagle{}), whereas \citet{Ghigna_et_al_2000} and \citet{Dolag_et_al_2009} report a power-law slope of $\alpha \approx -1$ in their $N$-body and lower-resolution hydrodynamical cluster simulations, respectively (the latter authors using the same subhalo finder as we do)\footnote{\citet{Ghigna_et_al_2000} quote a power-law slope of -2, but this is for the mass function defined as $\mathrm{d}n/\mathrm{d}M$, not $\mathrm{d}n/\mathrm{d}\log_{10}M$ as we show here.}.

The power-law slope is consistent between all three zones in our cluster simulations and the field, as determined from the original \eagle{} simulations (blue/purple). The normalization, on the other hand, clearly depends on environment, with a suppression of $\sim$0.7 dex within $\rvir$, and a very small deficiency ($\lesssim 0.1$ dex) even in the 5--10 $\rvir$ zone (orange). The former may partly reflect limitations in the \textsc{subfind} subhalo finder (e.g.~\citealt{Muldrew_et_al_2011}), but these authors show that beyond $\sim1.5 r_{200}$ \textsc{subfind} does accurately recover the total masses of subhaloes, so this is unlikely to significantly affect the outermost zone. Irrespective of this, we can conclude that the excess of massive galaxies in and around clusters is not linked to an excess of (massive) subhaloes.

\subsubsection{Stellar fractions of subhaloes}
\label{sec:msub_mstar}
\begin{figure}
  \centering
    \includegraphics[width=\columnwidth]{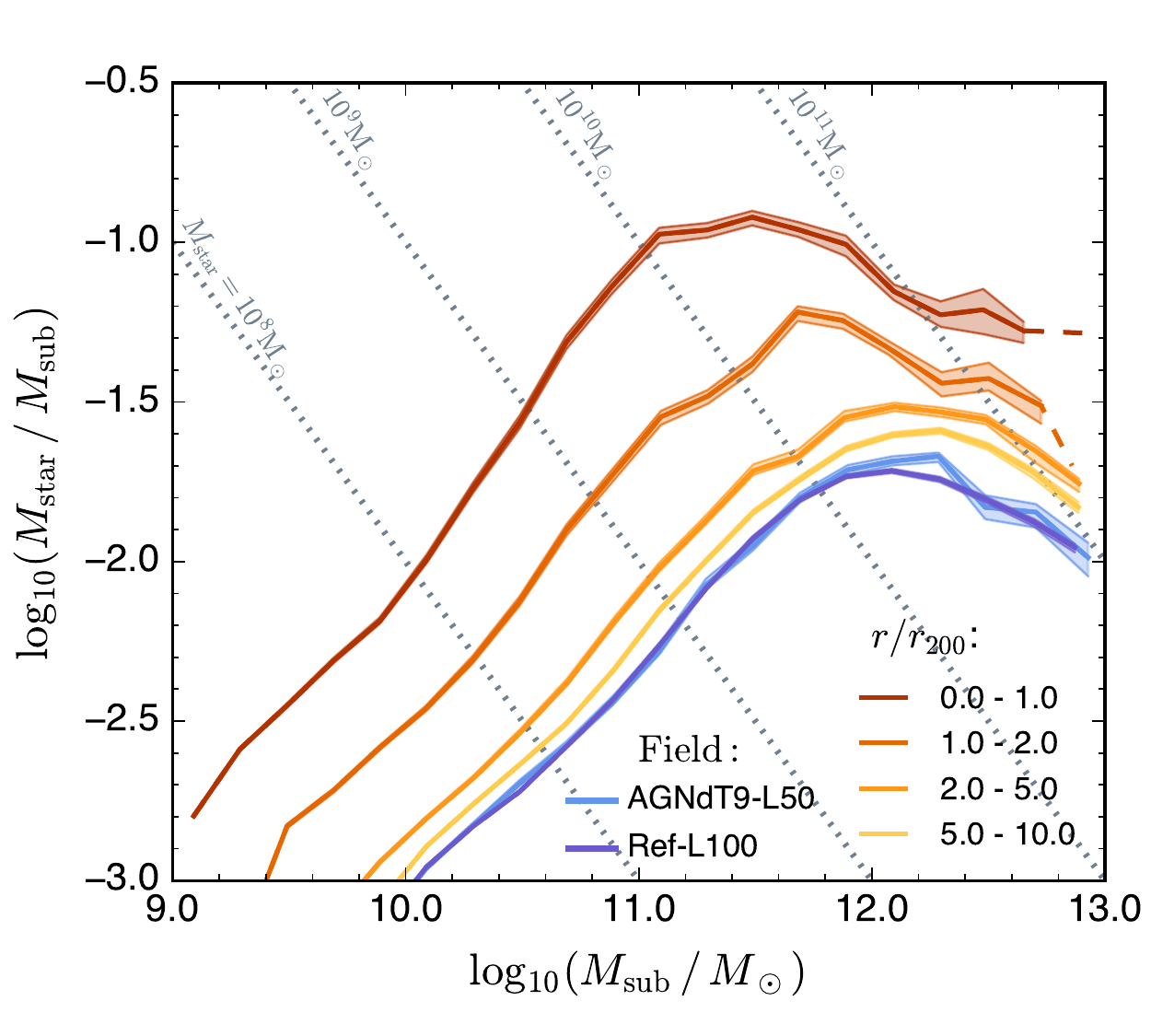}
       \caption{Median stellar mass fraction of Hydrangea $z=0$ subhaloes as a function of total subhalo mass $M_\text{sub}$, in different simulation zones (different colours). Shaded bands indicate $1\sigma$ uncertainties on the median. Dashed lines are used where there are fewer than 10 galaxies per 0.2 dex bin. Grey dotted lines indicate the location of galaxies with constant stellar mass as indicated near the top. There is a striking shift of the distribution towards lower subhalo masses at approximately constant stellar mass near the cluster centre (darker lines), but also an offset towards higher stellar fractions on the far cluster outskirts.}
    \label{fig:fstar_mtot}
\end{figure}

Our results above suggest that subhaloes (including centrals) in and around galaxy clusters have stellar mass fractions that differ from the field, which we confirm explicitly in Fig.~\ref{fig:fstar_mtot}. Field galaxies from the \eagle{} simulations (blue/purple) show an increasing stellar fraction at low (sub-)halo mass, with a peak at subhalo masses of $M_\text{sub} \approx 10^{12}\, \msun$ and subsequent decline towards higher masses; as \citet{Schaye_et_al_2015} have shown, this behaviour agrees quantitatively with what is inferred from observations within the framework of abundance matching (e.g.~\citealt{Behroozi_et_al_2013, Moster_et_al_2013}).

While cluster galaxies generally follow the same trend, there are two significant differences. At $r < 2\,\rvir$, stellar fractions are significantly higher at fixed $M_\text{sub}$ than in the field, especially at low subhalo masses (e.g.~a +1.4 dex offset at $10^{11} \msun$ inside $\rvir$), and the peak stellar fraction is shifted systematically to lower subhalo mass. Both of these differences are consistent with stripping of non-stellar mass (gas and dark matter), while the stellar mass remains constant, as indicated by the grey dotted lines in Fig.~\ref{fig:fstar_mtot}. To some extent, this may reflect artificial `stripping' by the \textsc{subfind} code which does not detect all bound particles as members of the subhalo, but with \citet{Muldrew_et_al_2011} reporting a detection efficiency of $\sim$50 per cent (0.3 dex) at $0.5\,\rvir$ (approximately the median radius of galaxies in the innermost bin), most of the stellar fraction difference is likely real.

At $r > 2\rvir$, stellar fractions remain higher than in the field, but the peak stellar fraction is located at approximately the same subhalo mass ($M_\text{sub} \approx 10^{12}\, \msun$), or plausibly shifts slightly \emph{higher} (by $\lesssim 0.2$ dex). This excess can therefore not be explained by halo stripping, and instead suggests that \emph{galaxy formation is more efficient near massive clusters}. A similar offset is seen when only considering central galaxies (not shown), so the offset is not due to differing fractions of satellites in different environments. We note that \citet{Moster_et_al_2013} found a scatter of only $\sim$0.15 dex in the $\mstar$--$M_\text{sub}$ relation, which is much less than the systematic offset with environment identified here ($\sim$0.3 dex in the 2--5 $\rvir$ bin, and even stronger at smaller radii).

\begin{figure}
  \centering
    \includegraphics[width=\columnwidth]{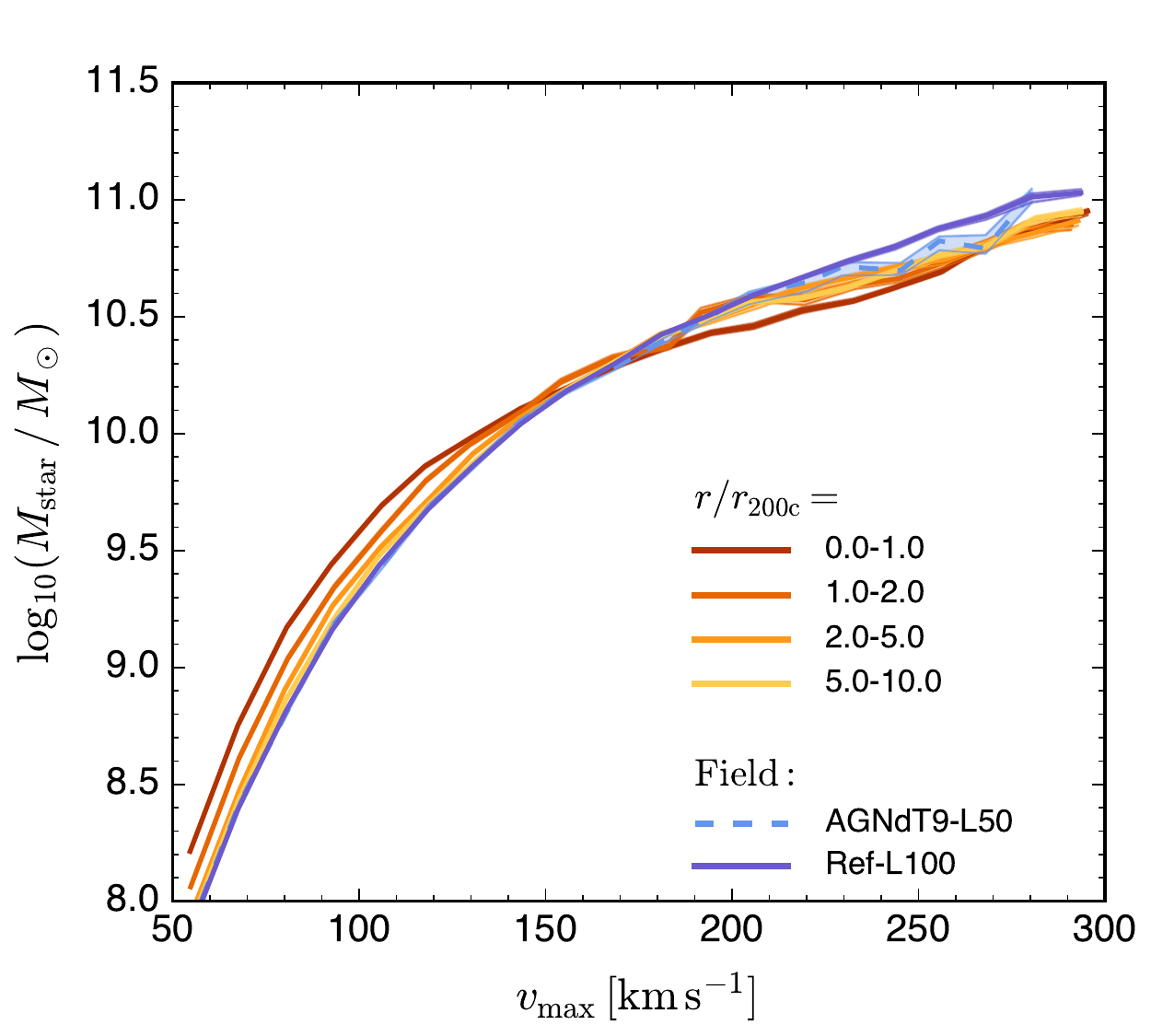}
       \caption{Stellar mass of Hydrangea galaxies as a function of maximum circular velocity, $v_\mathrm{max}$, compared to the field (blue/purple). In contrast to the comparison at fixed total mass presented in Fig.~\ref{fig:fstar_mtot}, there is no strong environmental offset for massive galaxies, with a small deficit of stellar mass in massive cluster galaxies compared to field galaxies of the same $v_\mathrm{max}$.}
    \label{fig:mstar_vmax}
\end{figure}

In contrast, Fig.~\ref{fig:mstar_vmax} demostrates that the stellar masses of galaxies at fixed maximum circular velocity ($v_\mathrm{max}$), i.e.~the Tully-Fisher relation \citep{Tully_Fisher_1977}, exhibit hardly any environmental variation in the Hydrangea simulations, at least at the massive end ($v_\mathrm{max} \gtrsim 150 \text{km s}^{-1}$). Furthermore, what little offset there is points in the opposite direction, i.e.~galaxies around clusters contain marginally \emph{less} stellar mass than in the field. This confirms previous findings that $v_\mathrm{max}$ is a better predictor of stellar mass than (sub-)halo mass (e.g.~\citealt{Conroy_et_al_2006, Reddick_et_al_2013, Chaves-Montero_et_al_2016}).

\subsection{Star formation histories}
\begin{figure}
  \centering
    \includegraphics[width=\columnwidth]{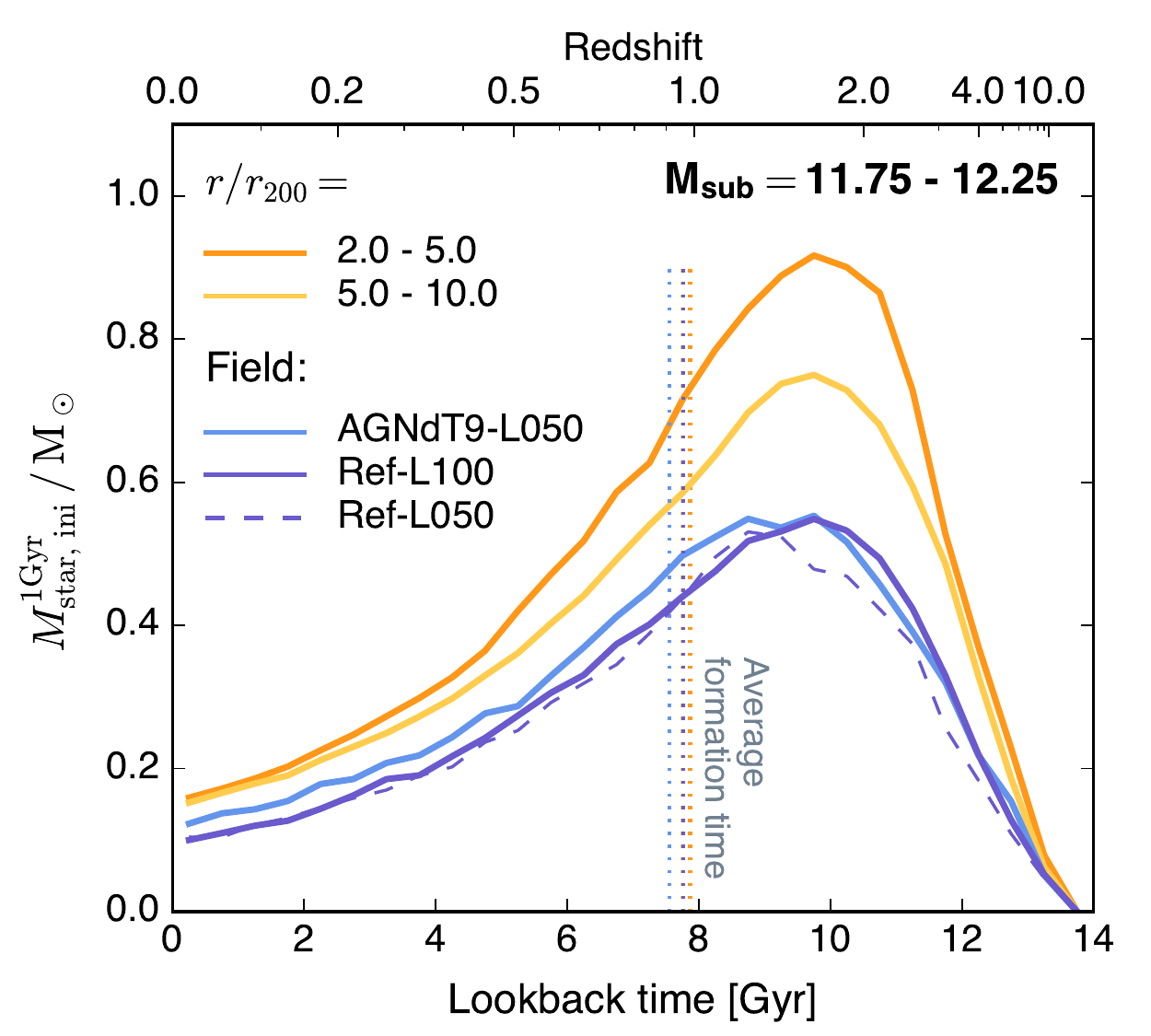}
       \caption{(Initial) stellar mass formed per Gyr as a function of lookback time in subhaloes of fixed total mass $M_\text{sub} = 10^{11.75}$--$10^{12.25} \msun$ at $z=0$ in different environments. Blue/purple lines show field galaxies from \eagle{}, while orange and yellow lines represent galaxies on the outskirts of Hydrangea clusters. The latter have had higher star formation throughout cosmic history, especially around $z = 2$. Vertical dotted lines show the corresponding mean formation redshifts, which are similar for all regions ($z_\text{av} \approx 1$).}
    \label{fig:halo_sfhist}
\end{figure}

To probe the predicted differences between star formation in clusters and the field in more detail, we have reconstructed the star formation history of our galaxies by separating their star particles into 27 narrow bins of formation lookback time (0--13.5 Gyr, i.e.~$\Delta t = 500$ Myr) and summing up the initial stellar mass in each bin\footnote{Note that this is not necessarily identical to the SFR history of the main progenitor, since our approach also includes stars accreted through mergers.}. The result is shown in Fig.~\ref{fig:halo_sfhist}, where we compare the star formation histories of galaxies within subhaloes of similar total mass at redshift $z = 0$ ($M_\text{sub} \approx 10^{12}\, \msun$) in the two outer cluster zones (2--5 $\rvir$ and 5--10 $\rvir$) of the Hydrangea simulations and in the \eagle{} periodic boxes (as `field'). Star formation has been more efficient near clusters than in the field throughout cosmic history, but particularly around the cosmic star formation rate peak at $z \approx 2$ (an excess of $\sim$80 per cent in the 2--5 $r_{200}$ zone compared to the field). We note that this does \emph{not} necessarily imply that star formation was more efficient at equal $z = 2$ subhalo mass, since (sub-)haloes near clusters are expected to have formed earlier \citep{Gao_et_al_2005} and will therefore have been more massive around the peak of star formation than subhaloes with the same $z = 0$ mass in the field. 

As exemplified in the bottom row of Fig.~\ref{fig:pretty}, the high-redshift progenitors of our clusters typically consist of a collection of similarly massive proto-cluster cores linked by dense filaments. It is therefore perhaps not too surprising that even those galaxies that did not collapse into the central cluster at $z=0$ still experienced a high-redshift evolution that differed significantly from average regions of the Universe. These differences leave detectable imprints in the properties of galaxies on the far outskirts of clusters, which are less subject to late-time processes such as star formation quenching. We will investigate the mechanisms impacting galaxies in high-redshift proto-cluster regions in detail in future work.

\subsection{Halo concentrations}
We have so far characterised subhaloes mainly by their total mass. However, for a galaxy like the Milky Way, the virial radius is $\sim$300 kpc, while star formation is restricted to the central $\sim$10 kpc. This opens the possibility that differences in the stellar-to-total mass ratio are caused by differences in halo concentrations: the more concentrated a halo is, the larger the fraction of its total mass (and plausibly also of its baryon content) that is compressed into the dense centre and can be turned into stars. In addition, the potential wells in more concentrated haloes are deeper so that feedback is less efficient, and because halo concentration correlates with formation time, such haloes will also have had more time to form stars. \citet{Matthee_et_al_2017} have shown that, in \eagle{}, at fixed halo mass galaxy stellar masses do indeed exhibit a positive correlation with the concentration of their host halo.

To test the hypothesis that this is the cause of the environmental trends we have identified above, we have computed the concentrations of FoF haloes in the Hydrangea simulations, as well as in \eagle{}, in the same way as described in Section \ref{sec:runs}. To distinguish between stellar mass differences as cause and as effect of varying concentrations, we have done this for both the hydrodynamic simulations and the corresponding DM-only runs, which are linked as described in Section \ref{sec:runs}. We did not compute concentrations for satellite subhaloes, because they do not have a well-defined virial radius, and hence focus here on central galaxies only. 

We then compute a `field-equivalent' concentration for each FoF halo in the hydrodynamic and DM-only Hydrangea simulations. For this, we select all haloes in the \eagle{} Ref-L100 and DMO-L100 simulation, respectively, whose $\log_{10} \mvir$ differs by $< 0.1$ from the Hydrangea halo under consideration. However, due to the steepness of the halo mass function, this sample of comparison \eagle{} haloes will typically be biased towards the low-mass end of the $\mvir$ selection range, and hence have a median concentration that is higher than that of the Hydrangea halo even in the absence of any real environmental differences. To mitigate this, we bin the \eagle{} haloes into ten narrow bins of $\Delta \log_{10} \mvir = 0.02\,\text{dex}$, and compute a median concentration weighted by the inverse number of haloes in each of these bins. Averaged over all Hydrangea haloes with $\mvir > 10^{11} \msun$, this weighting scheme results in a bias in $\mvir$ that is less than 0.01 dex.

\begin{figure}
\centering

	\includegraphics[width=\columnwidth]{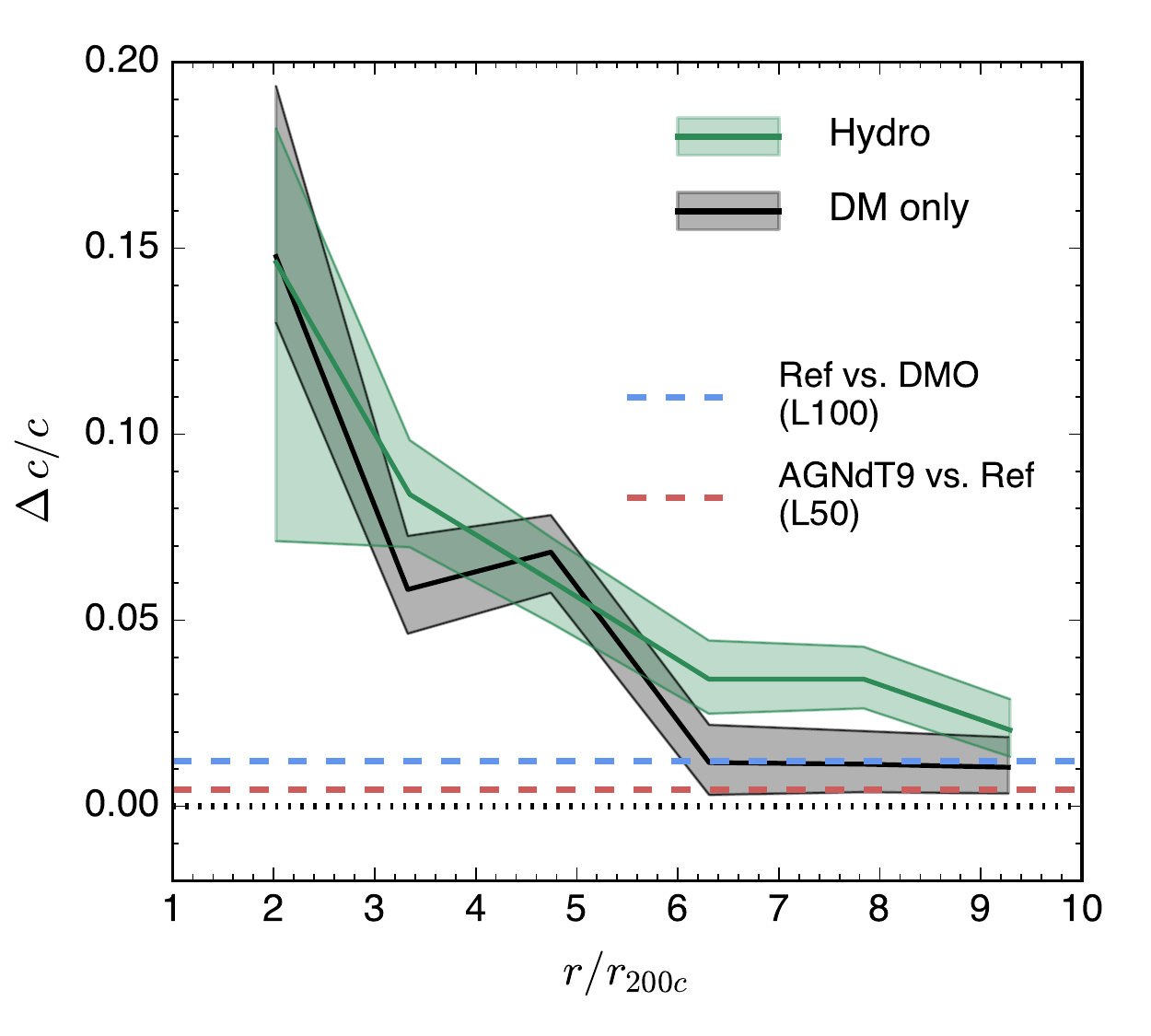}
	\caption{Relative difference in concentration of $\mvir \geq 10^{11.5}\, \msun$ haloes on the outskirts of Hydrangea clusters compared to the field, as a function of distance from the central cluster (see text for details). The green line compares hydrodynamic simulations, the black line the corresponding DM only runs. The two horizontal dashed lines indicate the difference in concentration between individually matched haloes in the \eagle{} L100 Ref vs.~DM only simulations (blue) and L50 AGNdT9 vs.~Ref runs (red). An environmental enhancement of halo concentration persists to at least 10 $\rvir$, is stronger in hydrodynamic than DM only simulations, and equals or exceeds concentration differences due to baryonic physics.}

\label{fig:mcplot_galaxy}
\end{figure}

Fig.~\ref{fig:mcplot_galaxy} shows the resulting concentration difference, $\Delta c/c \equiv (c_\text{Hydrangea}-c_\mathrm{\eagle{}})/c_\mathrm{\eagle{}}$ as a function of $r/\rvir$ for both the hydrodynamic (green) and DM-only Hydrangea simulations (black), for haloes with $\mvir > 10^{11.5}\,\msun$. In both cases, solid lines indicate running medians while shaded bands represent $1\sigma$ uncertainties. To compare the same haloes in both the DM-only and hydrodynamic simulations, we selected them based on $\mvir$ in the former, and then identified their counterparts in the latter via the links between their central subhaloes (see Section \ref{sec:runs}). 

Halo concentrations are clearly affected by the proximity to a cluster. The effect is strongest for haloes closest to the cluster, with an offset of 15 per cent at $r \approx 2 \rvir$. This environmental effect is significantly greater than concentration differences arising from the presence of baryons in the field: this can be tested by linking haloes in the \eagle{} Ref-L100 and DMO-L100 simulations in analogy to what we have done for Hydrangea, and only yields a concentration difference of $\sim$1 per cent in the halo mass range probed here (blue dashed line in Fig.~\ref{fig:mcplot_galaxy}). Even less significant are concentration differences arising from different parameterisations of AGN feedback, as we have tested by comparing the AGNdT9-L050 and Ref-L050 \eagle{} simulations (red dashed line).   

The environmental impact on halo concentrations decreases with increasing distance from the cluster, but does not reach zero even at the edge of our high-resolution region at $\sim$10 $\rvir$ (corresponding to distances of $\sim$20 Mpc). The difference between the hydrodynamic and DM-only simulations is small, which rules out (potentially uncertain) baryon effects as its dominant cause. 

A correlation between the concentration of dark matter haloes and their large-scale environment has already been demonstrated in DM-only simulations (e.g.~\citealt{Wechsler_et_al_2006, Gao_White_2007}) and is also present in \eagle{} \citep{Chaves-Montero_et_al_2016}. These studies analysed the dependence between concentration and the clustering of haloes of similar mass, and found that, at the low-mass end, the most concentrated haloes are more clustered than the least concentrated ones, but that this effect reverses at $\mvir \gtrsim 10^{13} \msun$. Our results demonstrate that a concentration increase persists to haloes in the vicinity of massive clusters, even on their far outskirts.

\subsubsection{Connection between halo concentration and stellar mass}
We have shown above that, at fixed halo mass, haloes near clusters contain more stellar mass and are more concentrated than in the field. We now test whether there is a connection between these two effects, by computing a `field-equivalent' stellar mass from the \eagle{} Ref-L100 simulation for each (central) Hydrangea galaxy, in analogy to the procedure for obtaining field-equivalent concentrations described above. As well as matching galaxies by $\mvir$ only, we have also repeated the procedure with a simultaneous match in $\mvir$ and $c$, requiring a maximum offset of 0.1 (0.05) in $\log_{10} \mvir$ ($\log_{10} c$) and computing weights as the product of the inverse number of galaxies in ten bins each in $\log_{10} \mvir$ and $\log_{10} c$. Because the concentrations, as well as halo masses, in the hydrodynamic simulations might themselves be affected by baryonic processes associated with the higher stellar mass content, we compute both quantities in the corresponding DM-only simulations and then link to the hydrodynamic runs as described in Section \ref{sec:runs} for the stellar masses.

\begin{figure}
  \centering
    \includegraphics[width=\columnwidth]{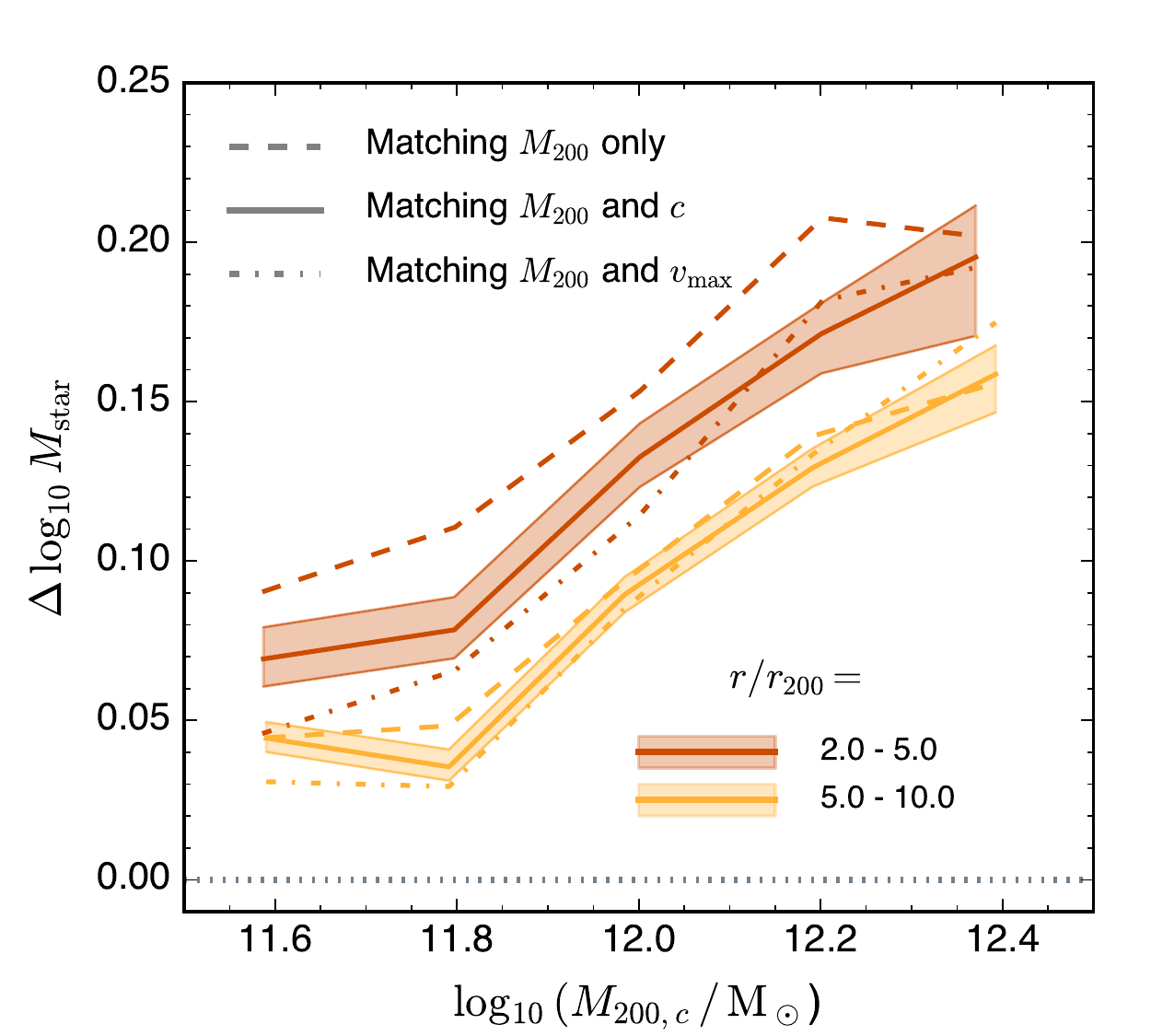}
       \caption{Excess stellar mass of central galaxies on the outskirts of clusters compared to the field. Galaxies are matched by the concentration and mass of their host halo (solid lines; shaded bands indicate statistical $1\sigma$ uncertainties on the running median). For comparison, the excess from matching only by mass is shown with dashed lines, and that from matching by mass and maximum circular velocity as dash-dot lines. The concentration difference (see Fig.~\ref{fig:mcplot_galaxy}) explains only a small part of the stellar mass excess around clusters, especially at $r > 5\rvir$.}

    \label{fig:delta_mstar}

\end{figure}

In Fig.~\ref{fig:delta_mstar}, we show the difference between the actual and field-equivalent stellar mass of Hydrangea galaxies, i.e. the effect of environment at fixed halo mass and concentration. Perhaps surprisingly, galaxies in both radial zones still show a significant environmental mass excess, which reaches $\sim$0.15 dex at $\mstar \approx 10^{12.4}\, \msun$. In fact, the excess is only marginally smaller than what is obtained without the additional match in concentration (dashed lines). The same is true when we match by the concentration as measured in the hydrodynamical, rather than DM-only, simulation (not shown). 

Evidently, higher halo concentrations are \emph{not} the cause of the stellar mass excess around clusters, and instead the two effects result from different physical processes associated with a galaxy's environment. Similarly, matching galaxies by $v_\text{max, DMO}$ instead of $c_\text{DMO}$ as second parameter (with a maximum offset of $\Delta \log_{10} v_\text{max} = 0.05$) achieves no significant reduction in the stellar mass offset (dash-dot lines in Fig.~\ref{fig:delta_mstar}). As we have shown in Fig.~\ref{fig:mstar_vmax}, the correlation between stellar mass and $v_\text{max}$ as measured directly in the hydrodynamic simulations is largely insensitive to environment. The fact that the same does not apply to $v_\text{max,\, DMO}$ indicates that the stellar mass offset near clusters is, in fact, not the result of differences between the DM haloes in the field and near clusters at fixed total mass, at least not to the extent that they are reflected in either their concentration or maximum circular velocity. Rather, the enhanced stellar mass and maximum circular velocity (including baryons) appear to be \emph{both} affected by an environmental effect that is predominantly, if not solely, due to baryons.  

Our results from Fig.~\ref{fig:delta_mstar} are qualitatively consistent with what was shown by \citet{Croton_et_al_2007} with semi-analytic models applied to the Millennium Simulation \citep{Springel_et_al_2005}: these authors found that the excess clustering signal for red galaxies is only marginally reduced when concentration is used in addition to halo mass to shuffle galaxies in their simulated catalogues. In principle, it is possible that the embedding into the ICM halo and its surrounding filaments exerts pressure on galaxies near clusters and thus prevents feedback-driven outflows from escaping. However, this effect is not captured by semi-analytic models, and furthermore \citet{Bahe_et_al_2012} showed that pressure confinement of satellite galaxies is generally ineffective. A more likely explanation is therefore that differences in the accretion history of field and cluster galaxies lead to stronger imprints in the present-day stellar mass fraction than in their halo concentrations.


\section{Summary and Discussion}
\label{sec:summary}

We have introduced the Hydrangea simulation suite, a set of 24 high-resolution cosmological hydrodynamical zoom-in simulations of massive galaxy clusters ($\mvir = 10^{14.0}-10^{15.4}\, \msun$) and their surroundings out to $10\,\rvir$ that form the key part of the C-EAGLE project. The simulations are run with the AGNdT9 galaxy formation model of the \eagle{} suite \citep{Schaye_et_al_2015}, and therefore allow a direct comparison between galaxy populations in the central regions of clusters, in their periphery, and in the field. They assume a Planck cosmology \citep{Planck_2014} and include sub-grid prescriptions for radiative cooling, reionization, star formation, metal enrichment, and energy feedback from both star formation and accreting supermassive black holes (see Section \ref{sec:eagle}). In this first paper, we have tested the stellar masses and star formation rates of our simulated galaxies, with the following main results:

\begin{enumerate}
\item Our simulations broadly reproduce the observed total stellar mass fraction in galaxy clusters and predict a slightly sub-linear scaling of stellar to halo mass. However, the stellar masses of simulated BCGs are too high by at least 0.2, and plausibly $> 0.5$, dex. The total and BCG stellar masses of clusters on the outskirts of an even more massive cluster follow the same relation as isolated clusters (Fig.~\ref{fig:mstar_r500}).

\item At $z \approx 0$, our simulations match several published satellite galaxy stellar mass functions (GSMF). The only mild discrepancy concerns low-mass galaxies ($\mstar < 10^{10} \msun$), which are predicted to be somewhat less numerous than in the SDSS analysis of \citet{Yang_et_al_2009} (Fig.~\ref{fig:gsmf_obscomp}). 

\item In qualitative agreement with observations, simulated cluster satellite galaxies have a quenched fraction (sSFR $\equiv$ SFR/$\mstar < 10^{-11}$ yr$^{-1}$) that is higher than for centrals with the same stellar mass. The quenched fraction excess at $\mstar \gtrsim 10^{10}\, \msun$ is close to the observed value ($\sim$60 per cent at $\sim 10^{10}\, \msun$). However, there are also quantitative discrepancies. The quenched fractions of both centrals and satellites are lower than observed at $\mstar \gtrsim 10^{10}\, \msun$, reaching only $\sim$70 per cent at $\mstar \approx 10^{11}\, \msun$ instead of near unity. At $\mstar \lesssim 10^{10}\, \msun$, the quenched fraction of satellites is too high and shows an artificial increase towards lower stellar masses, plausibly as a consequence of insufficient resolution (Fig.~\ref{fig:fquenched_obscomp}).

\item Normalized to total mass, the GSMF in our cluster simulations shows two subtle but significant differences from the field: a deficiency of low-mass ($\mstar \lesssim 10^{10} \msun$) galaxies within $\rvir$, and an excess of massive galaxies ($\mstar \gtrsim 10^{10} \msun$) from the centre to the far outskirts ($\sim$$10\rvir$). Neither of these effects depends significantly on cluster mass (Fig.~\ref{fig:gsmf_bias}).

\item The deficiency of low-mass galaxies is not primarily caused by tidal stripping, but emerges as a consequence of star formation quenching: it is only present in young stars (formed at $z < 1$), while the abundance of old stars in low-mass galaxies is consistent between clusters and the field (Fig.~\ref{fig:gsmf_bias_z1}).

\item The excess of massive galaxies is not caused by an excess of massive subhaloes on the outskirts of our simulated clusters (Fig.~\ref{fig:subhalo_bias}), and instead originates from a significantly higher ($\gtrsim$ 0.2 dex) stellar fraction at a given subhalo mass (Fig.~\ref{fig:fstar_mtot}). This is found to be due to higher levels of star formation in (proto-)cluster environments than in the field, especially at redshift $z \gtrsim 1$, with an excess star formation of up to 80 per cent in subhaloes with $M_\text{sub} \approx 10^{12}\, \msun$ compared to subhaloes with the same mass at $z=0$ in the field (Fig.~\ref{fig:halo_sfhist}).

\item At fixed mass, haloes near a cluster are more concentrated than in the field, out to $\sim$$10\rvir$ from the cluster centre (Fig.~\ref{fig:mcplot_galaxy}). However, this does not explain the higher stellar mass fractions around clusters, because a similarly high stellar mass excess still remains between haloes of similar mass \emph{and} concentration (Fig.~\ref{fig:delta_mstar}).

\end{enumerate}

The analysis presented here adds to the growing body of evidence that galaxy formation even far away from the centres of massive haloes is affected by the environment (e.g.~\citealt{Wetzel_et_al_2012, Lu_et_al_2012, Bahe_et_al_2013}). So far, large-scale environmental influence has been studied mainly in the context of the gas content and star formation rate of galaxies. According to our simulations, environment also affects the stellar masses of galaxies out to large radii, which is important because stellar mass is commonly used as the label to compare `similar' galaxies in the field and in dense environments. This may lead to unexpected complications in the interpretation of observational results if relevant physical processes, such as ram pressure stripping, do not only depend significantly on stellar mass, but also on e.g. the halo mass. A comparison of galaxies in different environments matched only by $\mstar$ may then be fundamentally biased. In future work, we will explore the consequences of this bias in more detail.

It is also important to keep in mind that our simulation model was not calibrated in any way to produce realistic environmental effects on galaxies. As discussed by \citet{Schaye_et_al_2015} and \citet{Crain_et_al_2015}, calibration of the \eagle{} model primarily involved the stellar masses and sizes of the overall galaxy population (i.e.~mostly centrals), while the modifications to the AGN subgrid model in AGNdT9 compared to Ref were motivated by hot gas fractions in groups that were higher than observed.  In light of this, the approximately realistic prediction of the quenched fraction excess, the stellar mass function, and total stellar mass in massive clusters is encouraging. Moreover, we demonstrate in a companion paper (Barnes et al., in prep.) that the hot gaseous haloes of our simulated clusters show approximately realistic global properties, such as hot gas fractions and X-ray emission, albeit with discrepancies in detail.

In the future, we will exploit this potential of the Hydrangea simulations to understand how the formation of galaxies in and around massive clusters differs from that of isolated galaxies, in terms of e.g.~their gas accretion, star formation activity, and morphological evolution.     

\section*{Acknowledgments}
We thank Lydia Heck for expert computational support with the Cosma machine in Durham, which was used for part of the work presented here. RAC is a Royal Society University Research Fellow. DJB and STK acknowledge support from STFC through grant ST/L000768/1. CDV acknowledges financial support from the Spanish Ministry of Economy and Competitiveness (MINECO) under the 2011 and 2015 Severo Ochoa Programs SEV-2011-0187 and SEV-2015-0548, and grants AYA2014-58308 and RYC-2015-18078. PAT (ORCID 0000-0001-6888-6483) acknowledges support from the Science and Technology Facilities Council (grant number ST/L000652/1). The Hydrangea simulations were in part performed on the German federal maximum performance computer~``HazelHen'' at the maximum performance computing centre Stuttgart (HLRS), under project GCS-HYDA / ID 44067 financed through the large-scale project~``Hydrangea'' of the Gauss Center for Supercomputing. Further simulations were performed at the Max Planck Computing and Data Facility in Garching, Germany. This work also used the DiRAC Data Centric system at Durham University, operated by the Institute for Computational Cosmology on behalf of the STFC DiRAC HPC Facility (www.dirac.ac.uk). This equipment was funded by BIS National E-infrastructure capital grant ST/K00042X/1, STFC capital grant ST/H008519/1, and STFC DiRAC Operations grant ST/K003267/1 and Durham University. DiRAC is part of the National E-Infrastructure. We also gratefully acknowledge PRACE for awarding the EAGLE project access to the Curie facility based in France at Tr\`{e}s Grand Centre de Calcul. Support was also received via the Interuniversity Attraction Poles Programme initiated by the Belgian Science Policy Office ([AP P7/08 CHARM]), the National Science Foundation under Grant No. NSF PHY11-25915, and the UK Science and Technology Facilities Council (grant numbers ST/F001166/1 and ST/I000976/1) via rolling and consolidated grants awarded to the ICC. The research was supported by the Netherlands Organisation for Scientific Research (NWO), through VICI grant 639.043.409, and by the European Research Council under the European Union's Seventh Framework Programme (FP7/2007-2013) / ERC Grant agreement 278594-GasAroundGalaxies. This research has made use of NASA's Astrophysics Data System. All figures in this paper were produced using the \textsc{Astropy} \citep{Astropy_2013} and \textsc{Matplotlib} \citep{Hunter_2007} Python packages.
\bibliographystyle{mnras}
\bibliography{bibliography}


\begin{appendix}

\section{Summary of Hydrangea clusters}
\label{app:summary}
In Table \ref{tab:info}, we provide information about each of the 30 C-EAGLE clusters at $z=0$. Masses are computed as the total mass within spherical apertures centred on the potential minimum of the cluster within which the average density equals 200 (500) times the critical, as well as 200 times the mean, density of the Universe. Concentrations are obtained as described in Section \ref{sec:runs}, by fitting an NFW profile to the dark matter density profile between 0.05 and 1 $\rvir$, following \citet{Neto_et_al_2007} and \citet{Schaller_et_al_2015}. The position coordinates $x$, $y$, and $z$ (in units of pMpc) specify the centre of each re-simulation region in the original parent simulation (see Barnes et al., in prep., and Appendix \ref{app:ics} for a description of how our high-resolution initial conditions were generated). The dominance measure ($D_5$) specifies the distance (in pMpc) from the central cluster to the nearest halo with $\mvir$ at least 1/5 of the central cluster. $D_5$ is calculated from the parent dark matter only simulation, because not all zoom regions contain such a massive secondary halo within their high-resolution region. Finally, we give the number of galaxies with $\mstar \geq 10^9\,\msun$ within 1 and 10 $\rvir$ (5 $\rvir$ for the six simulations that are not part of Hydrangea) from the potential minimum of the central cluster.

\begin{table*}
\caption{Overview of the 30 C-EAGLE simulations at redshift $z=0$. The 24 Hydrangea simulations with high-resolution regions extending to at least 10 $\rvir$ from the cluster centre are listed first. The last six entries, below the horizontal line, represent the six additional haloes simulated only out to 5 $\rvir$. We provide the radii within which the average density equals 200 (500) times the critical, and 200 times the mean, density; the total mass enclosed in these radii, as well as the stellar mass within $r_{500c}$; the centre of the zoom-in region in the (3200 pMpc)$^3$ parent simulation; the best-fit NFW concentration of the central cluster halo; the dominance parameter $D_5$, defined as the distance to the nearest halo whose mass is at least one fifth of that of the central cluster (determined from the parent simulation); and the number of galaxies with $M_\text{star, 30pkpc} \geq 10^9 \msun$ within 1 and 10 $\rvir$ from the potential minimum of the cluster. X-ray properties of the central clusters at $z=0.1$ are provided in the companion paper by Barnes et al. (in prep.). $^\dag$For the six clusters simulated only to $5\rvir$, the last column instead gives the number of galaxies within this radius.}

\begin{tabular}{crrrrrrrrrrrrrr} \hline \hline
Halo & $r_{200c}$ & $r_{200m}$ & $r_{500c}$ & $M_{200c}$ & $M_{200m}$ & $M_{500c}$ & $M_\text{500c}^\text{star}$  & $x$ & $y$ & $z$ & $c_\text{NFW}$ & $D_5$ & $N_\text{galaxies}$ \\
ID &  \multicolumn{3}{c}{[pMpc]} & \multicolumn{4}{c}{[$\log_{10} (M/\msun)$]} & \multicolumn{3}{c}{[pMpc]} & & [pMpc] & $< r_{200c}$ & $< 10 r_{200c}$ \\
\hline
CE-0 & 1.03 & 1.74 & 0.68 & 14.07 & 14.24 & 13.92 & 12.21 & 313.65 & 2218.64 & 2652.71 & 5.3 & 11.4 & 36 & 181 \\
CE-1 & 1.02 & 1.63 & 0.65 & 14.05 & 14.15 & 13.87 & 12.16 & 2598.97 & 2552.80 & 2266.29 & 3.7 & 6.9 & 34 & 163 \\
CE-2 & 1.02 & 1.63 & 0.65 & 14.05 & 14.15 & 13.87 & 12.16 & 2889.69 & 2880.09 & 355.44 & 6.1 & 15.8 & 34 & 163 \\
CE-3 & 1.09 & 1.84 & 0.70 & 14.14 & 14.31 & 13.97 & 12.15 & 2608.58 & 2831.41 & 908.38 & 6.5 & 9.9 & 49 & 243 \\
CE-4 & 1.17 & 1.89 & 0.78 & 14.23 & 14.34 & 14.10 & 12.29 & 1720.84 & 2253.49 & 2670.52 & 4.2 & 5.2 & 68 & 322 \\
CE-5 & 1.09 & 1.90 & 0.72 & 14.15 & 14.35 & 13.99 & 12.29 & 583.22 & 908.50 & 1669.79 & 6.7 & 14.2 & 42 & 294 \\
CE-6 & 1.27 & 2.16 & 0.81 & 14.34 & 14.52 & 14.15 & 12.35 & 2624.03 & 2241.14 & 304.69 & 3.6 & 17.6 & 76 & 380 \\
CE-7 & 1.27 & 2.17 & 0.81 & 14.34 & 14.53 & 14.16 & 12.37 & 1272.32 & 2452.95 & 1288.05 & 4.6 & 7.5 & 76 & 452 \\
CE-8 & 1.23 & 2.12 & 0.79 & 14.30 & 14.49 & 14.12 & 12.36 & 486.08 & 735.81 & 357.66 & 4.5 & 16.2 & 67 & 338 \\
CE-9 & 1.39 & 2.36 & 0.92 & 14.46 & 14.63 & 14.32 & 12.48 & 1368.63 & 1452.69 & 2207.20 & 5.2 & 9.1 & 84 & 486 \\
CE-10 & 1.29 & 2.21 & 0.82 & 14.36 & 14.55 & 14.17 & 12.45 & 2616.89 & 1602.52 & 1876.43 & 4.8 & 10.3 & 90 & 446 \\
CE-11 & 1.43 & 2.34 & 0.94 & 14.49 & 14.63 & 14.35 & 12.51 & 2564.49 & 678.34 & 1356.74 & 6.5 & 8.8 & 109 & 537 \\
CE-12 & 1.55 & 2.49 & 1.03 & 14.60 & 14.71 & 14.47 & 12.71 & 1165.85 & 1386.20 & 1010.20 & 4.7 & 26.5 & 148 & 506 \\
CE-13 & 1.57 & 2.52 & 1.07 & 14.61 & 14.72 & 14.51 & 12.63 & 998.80 & 1511.46 & 1963.65 & 6.3 & 11.4 & 131 & 498 \\
CE-14 & 1.62 & 2.66 & 0.98 & 14.66 & 14.79 & 14.41 & 12.52 & 276.94 & 1459.94 & 2042.48 & 2.5 & 10.8 & 179 & 734 \\
CE-15 & 1.71 & 2.73 & 1.05 & 14.73 & 14.83 & 14.49 & 12.74 & 2015.45 & 737.45 & 1738.86 & 2.2 & 6.4 & 203 & 957 \\
CE-16 & 1.74 & 2.84 & 1.17 & 14.75 & 14.88 & 14.63 & 12.76 & 717.52 & 2244.68 & 609.33 & 7.0 & 9.2 & 202 & 1179 \\
CE-18 & 1.87 & 3.03 & 1.23 & 14.84 & 14.96 & 14.70 & 12.64 & 793.71 & 864.02 & 1612.59 & 4.8 & 27.0 & 261 & 1061 \\
CE-21 & 1.99 & 3.34 & 1.24 & 14.93 & 15.09 & 14.71 & 12.87 & 1139.47 & 909.91 & 948.80 & 3.3 & 11.9 & 306 & 1901 \\
CE-22 & 2.14 & 3.72 & 1.39 & 15.02 & 15.23 & 14.86 & 12.85 & 2078.36 & 2319.21 & 843.85 & 4.4 & 5.2 & 362 & 3153 \\
CE-24 & 2.27 & 3.61 & 1.52 & 15.09 & 15.19 & 14.97 & 12.82 & 306.88 & 996.23 & 2870.46 & 5.0 & 21.9 & 425 & 1701 \\
CE-25 & 2.36 & 3.87 & 1.47 & 15.15 & 15.28 & 14.93 & 12.91 & 1028.05 & 1272.37 & 1276.27 & 2.5 & 20.6 & 497 & 2185 \\
CE-28 & 2.50 & 4.06 & 1.68 & 15.22 & 15.34 & 15.10 & 13.02 & 1390.16 & 1049.82 & 2040.15 & 3.7 & 16.2 & 556 & 2804 \\
CE-29 & 2.82 & 4.61 & 1.61 & 15.38 & 15.51 & 15.04 & 12.96 & 1070.13 & 2140.38 & 1498.16 & 1.8 & 30.1 & 826 & 3788 \\ \hline
CE-17 & 1.65 & 2.74 & 1.02 & 14.68 & 14.83 & 14.45 & 13.07 & 216.56 & 1847.43 & 2889.33 & 2.7 & 14.5 & 180 & 381$^\dag$ \\
CE-19 & 1.86 & 3.07 & 1.21 & 14.84 & 14.98 & 14.68 & 13.13 & 805.68 & 319.03 & 1136.84 & 3.4 & 9.0 & 291 & 704$^\dag$ \\
CE-20 & 1.77 & 2.87 & 1.16 & 14.78 & 14.89 & 14.62 & 13.15 & 2693.84 & 1783.70 & 2955.12 & 5.0 & 14.4 & 216 & 449$^\dag$ \\
CE-23 & 1.99 & 3.34 & 1.31 & 14.92 & 15.09 & 14.77 & 12.92 & 2033.86 & 2989.23 & 2715.06 & 3.1 & 10.0 & 314 & 848$^\dag$ \\
CE-26 & 2.39 & 3.89 & 1.56 & 15.16 & 15.29 & 15.00 & 13.23 & 2818.50 & 1262.96 & 1993.58 & 5.5 & 11.6 & 468 & 1083$^\dag$ \\
CE-27 & 2.39 & 3.82 & 1.64 & 15.16 & 15.26 & 15.07 & 13.16 & 2646.97 & 913.51 & 2629.65 & 7.2 & 20.2 & 252 & 475$^\dag$ \\ \hline

\end{tabular}

\label{tab:info}
\end{table*}

\section{Generation of initial conditions}
\label{app:ics}

Based on the 3.2 cGpc parent simulation \citep{Barnes_et_al_2017}, zoomed initial conditions (ICs) for our cluster re-simulations were generated with the second-order Lagrangian perturbation theory code \textsc{ic\_2lpt\_gen} \citep{Jenkins_2010} and using the public \textsc{Panphasia} white noise field \citep{Jenkins_2013}\footnote{The phase descriptor of the parent simulation is [Panph1, L14, (2152, 5744, 757), S3, CH1814785143, EAGLE\_L3200\_VOL1].}. This approach is similar to what was done by \citet{Barnes_et_al_2017} for the MACSIS project and is described in more detail in the companion paper by Barnes et al. (in prep.). As described in Section \ref{sec:large_motivation}, we required that a sphere of radius $10\,\rvir$ around each cluster centre -- defined as the potential minimum of the cluster halo -- be free from low-resolution boundary particles at redshift $z=0$. Within this high-resolution region, dark matter particle masses are nearly the same\footnote{The particle masses realized by our zoom-in ICs generator cannot be specified to arbitrary precision, as they are formed from $10^3$ glass tiles that have to be accommodated within the masked region. The actual particle masses therefore vary slightly between different zoom simulations, by $< 3$ per cent.} as in the `intermediate' resolution runs of the \eagle{} suite, i.e.~$m_\text{dm} \approx 9.7 \times 10^6\, \msun$. From these dark matter only ICs, the ICs including baryons were derived as described in appendix B2 of \citet{Schaye_et_al_2015}: each original particle is split into one dark matter and one SPH (gas) particle, with a mass ratio of $\Omega_\text{baryon}/(\Omega_\text{matter}-\Omega_\text{baryon}) = 0.186$. The initial baryon particle mass in our simulations is therefore $m_\text{baryon} \approx 1.81 \times 10^6\, \msun$. 

As a technical detail, we note that the particle indexing in C-EAGLE (including Hydrangea) is different from \eagle{}. In the latter, the particle IDs in the original DM-only ICs encode the particle's position along the Peano-Hilbert curve (see appendix B3 of \citealt{Schaye_et_al_2015}). While this makes it easy to link each particle to its initial position, it leads to very large ID numbers that cannot easily be used as keys to compare particles between different outputs. In C-EAGLE (including Hydrangea), the original DM-only IDs are therefore assigned in running order from 1 to $N_\text{part}$. As in \eagle{}, when we create the full hydrodynamic ICs, the ID of the dark matter particle is assigned to be exactly twice that of the original particle, and that of the gas particle one more than its corresponding dark matter particle; thus all DM particles have even, and all baryon particles odd, ID numbers.

\section{Tracing of subhaloes between outputs}
\label{app:tracing}
To fully utilise the information provided by our simulations, it is necessary to be able to link galaxies between outputs to reconstruct their individual formation histories. Although the results presented in this paper do not rely significantly on this ability, we will exploit this information in future work. For reference, we therefore describe here our subhalo tracing method, which is adapted from \citet{Bahe_McCarthy_2015}. We will in this context use the term `galaxy' to refer to the physical entity that is present in multiple snapshots (irrespective of whether its stellar mass is zero or not), and the term `subhalo' for each individual identification of a galaxy in one snapshot. 

Our tracing procedure exploits the ability to identify individual dark matter particles in different snapshots through their unique particle IDs. As a first step, we link in each pair of neighbouring snapshots any two subhaloes that share at least 20 dark matter particles, as long as these particles represent at least one per cent of all DM particles in the lower-redshift snapshot. We note that in \citet{Bahe_McCarthy_2015}, we had also included star particles to allow tracing galaxies beyond the point of disruption of their dark matter halo. This is not done here, because the improved resolution of the Hydrangea simulations means that even subhaloes with a dark matter mass of only $\sim 2\times 10^8\, \msun$ are resolved by 20 DM particles. 

In the simplest possible scenario, each subhalo in a given snapshot $i$ would `receive' only one link from a subhalo in the preceeding snapshot ($i-1$) , and `send' one link to a subhalo in the subsequent snapshot ($i+1$). In this case, we could unambiguously identify these subhaloes as representing the same galaxy in all three snapshots.

In reality, however, galaxies are expected to exchange particles between each other (e.g.~in mergers), so that one subhalo identified in snapshot $i$ will, in general, be linked to multiple others in $i+1$ (and vice versa). As a second step, we therefore have to select the best-matching links between $i$ and $i+1$. For this purpose we rank all links sent from a given subhalo in $i$, and all those received by a given subhalo in $i+1$, by their total mass -- i.e. the sum of the particle masses contributing to this link, which in our DM-only case is equivalent to the number of particles. In this way, each link is assigned a `sender rank' and a `receiver rank'. We then select those links with the highest receiver rank at each subhalo in $i+1$. If one subhalo in $i$ sends multiple links with equal receiver ranks, only the one with the highest sender rank is considered out of these.

In practice, the majority of selected links are those with the highest receiver rank, i.e.~those contributing the largest amount of DM particles to a given subhalo in snapshot $i+1$. Under certain circumstances, it may however be appropriate to select a link with a lower receiver rank, in particular if multiple links received and sent by one subhalo have comparable masses (e.g.~in complex mergers). After selecting the receiver-rank 0 links, we therefore then iterate through the nine next-highest receiver ranks at each subhalo which have a mass of at least two thirds of the highest rank link (if they exist), and select from those in analogy with the rank-0 selection described above.

The reason for this double-ranking (by sender and receiver) is that it prevents situations where a small subhalo accreted onto a more massive one is misidentified as the latter's progenitor, while allowing subhaloes that lose the majority of their mass due to, for example, tidal stripping, to be traced for as long as possible. We repeat this process for each pair of neighbouring snapshots to obtain a continuous history of all galaxies in our simulation. In each snapshot, any subhalo that has no receiving link selected is assumed to represent a newly formed galaxy.

\begin{figure}
  \centering
    \includegraphics[width=\columnwidth]{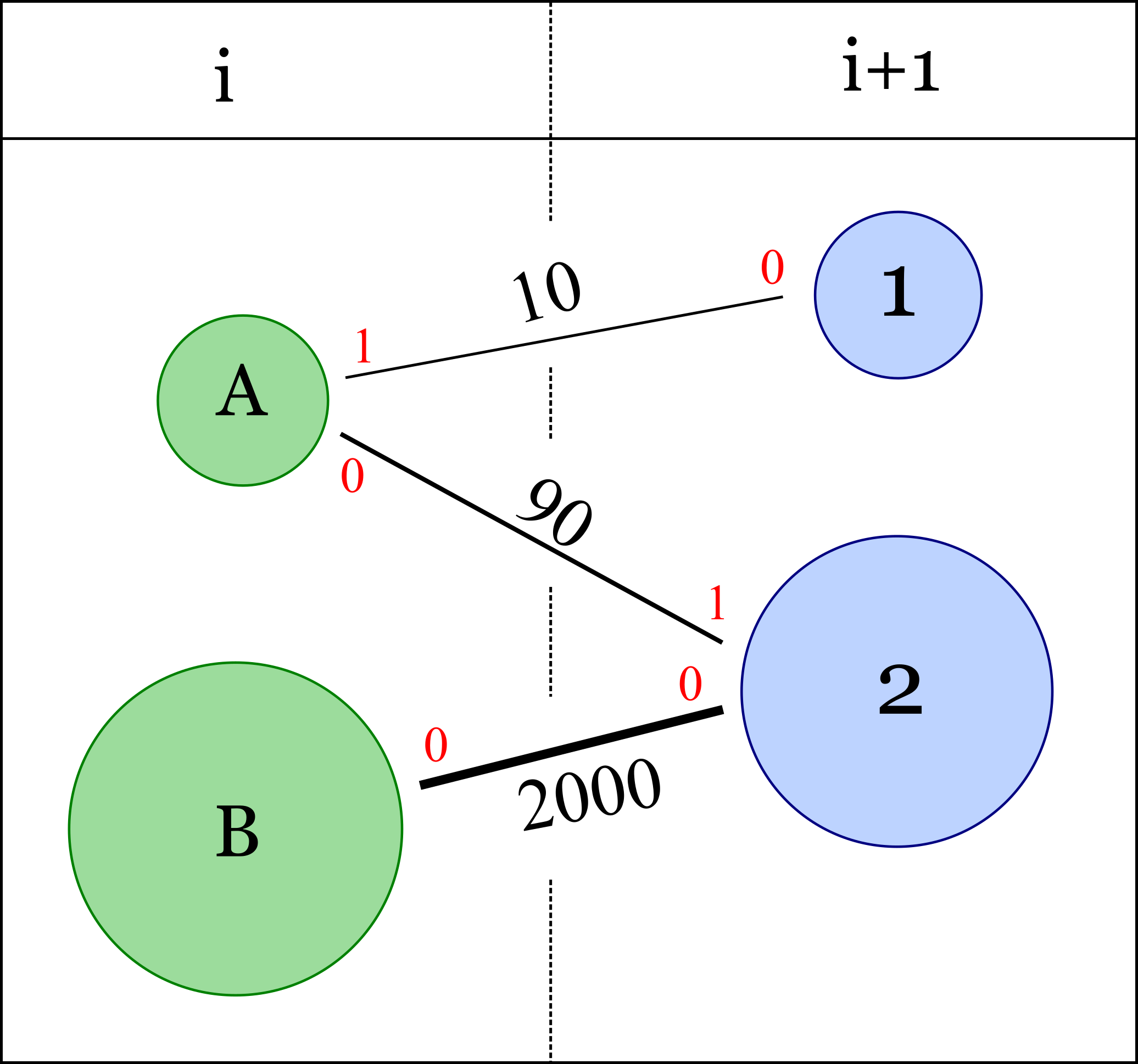}    
       \caption{Schematic example of the links between four subhaloes in two consecutive snapshots. As explained in the text, our tracing algorithm ensures that subhaloes A and 1 are linked into one galaxy, rather than being treated as merged onto 2.}
      \label{fig:tracing}
\end{figure}

As an illustration of this linking procedure, consider the situation depicted in Fig.~\ref{fig:tracing}: two subhaloes each in consecutive snapshots ($i$, $i+1$) are connected by three links with 10, 90, and 2000 particles, respectively. It is unambiguous that subhaloes B and 2 represent the same galaxy, since they are each other's best-matching progenitor \emph{and} descendant. Subhalo A, on the other hand, could be treated as either having merged with 2, or as representing the same galaxy as 1, but with most of its matter transferred onto subhalo 2 (e.g.~through tidal stripping). We prefer the second option, since it maximizes the time for which a galaxy orbiting in a cluster can be tracked.  

An additional complication is that subhalo finders such as \textsc{subfind} are known to have difficulty identifying subhaloes in dense backgrounds, such as the central regions of a galaxy cluster (e.g.~\citealt{Muldrew_et_al_2011}). Unaccounted for, this would lead to spurious subhalo ``disruption'' (when a subhalo still physically exists, but is not identified as such) and ``formation'' (if it is re-identified later). To mitigate this, we also trace subhaloes over two consecutive snapshot intervals by forming what we call `long links' between each pair of snapshots separated by one snapshot between them, in analogy to the `short links' described above. In the simplest case, the temporary non-identification will leave a subhalo A in the first snapshot $i$ without a (short-link) descendant, and a counterpart B in the second snapshot ($k \equiv i+2$) without a (short-link) progenitor. Provided A and B are connected by a long link, we can then join them together from $i$ to $k$ and skip the missing identification in snapshot $j$ in-between.  

However, it is also possible that between redshifts $i$ and $k$, the subhalo accretes another, smaller subhalo, which would then be identified as its progenitor although physically it is not (c.f.~above). We therefore also allow selection of long links between subhaloes that already have an identified (short-link) progenitor or descendant in the immediately neighbouring snapshot, provided this results in a better match of particles between subhaloes. 

The procedure described above allows our code to robustly follow self-bound structures through time, accounting for subhalo formation, merging and disruption, as well as temporary non-identification of subhaloes in dense environments. In the future, we also intend to run alternative substructure identification and subhalo tracing codes on our galaxy cluster simulations, and to compare the results.

\end{appendix}

\end{document}